\documentclass[a4paper, amsfonts, amssymb, amsmath, aps, pra, reprint, showkeys, nofootinbib]{revtex4-2}
\usepackage[english]{babel}
\usepackage[utf8]{inputenc}
\usepackage{graphicx}
\usepackage[left=18mm,right=18mm,top=35mm,columnsep=15pt]{geometry} 
\usepackage[T1]{fontenc}
\usepackage[pdftex, pdftitle={Enhanced Purcell factor for nanoantennas supporting interfering resonances}, pdfauthor={SB}]{hyperref}
\begin{document}
\title{Enhanced Purcell factor for nanoantennas supporting interfering resonances}

\author{
R\'{e}mi Colom$^1$, 
Felix Binkowski$^1$, 
Fridtjof Betz$^1$,
Yuri Kivshar$^{2}$,
Sven Burger$^{1,3}$}

\email[Correspondence email address: ]{burger@zib.de}
\affiliation{$^1$Zuse Institute Berlin, Takustraße 7, 14195 Berlin, Germany\\
$^2$Nonlinear Physics Center, Research School of Physics, 
Australian National University, Canberra ACT 2601, Australia\\
$^3$JCMwave GmbH, Bolivarallee 22, 14050 Berlin, Germany}

\begin{abstract}
We study the effect of coupled resonances and quasi-bound states in the continuum (quasi-BICs) on the Purcell factor in dielectric resonant nanoantennas. We analyze numerically interfering resonances in a nanodisk {\it with} and {\it without} a substrate when the modes are coupled to an emitter localized inside the nanodisk, and we quantify the modal contributions to the Purcell factor also reconstructing the radiation patterns of the resonant system. It is revealed that the Purcell effect can be boosted substantially for a strong coupling of resonances in the quasi-BIC regime.  
\end{abstract}

\maketitle

\section{Introduction}

Resonances play a central role in the control of light-matter interactions in nanophotonics.  Plasmonic resonances enable such a control via large near-field enhancements \cite{maier2005,maier2007}, which allows, e.g.,
for realizing plasmonic nanoantennas to tailor the radiation from quantum emitters~\cite{novotny2011,novotny2012}. Recently, the excitation of Mie-type resonant modes~\cite{Evlyukhin_2010,Garcia-Etxarri_2011} in high-refractive-index 
dielectric resonators has proven to be very useful for a wide range of applications, from the enhancement of nonlinear effects to a resonant control of the phase in metasurfaces~\cite{kuznetsov2016,Barreda_AIP_2019}.  
One important figure of merit for measuring the effect of resonances on light-matter interactions is their quality factor ({\it $Q$-factor}), that quantifies the ability of a structure to trap light and to enhance the electromagnetic fields.
Nanoresonators act as nanoantennas for strongly localized light sources, like quantum dots or defects in crystalline lattices, which can allow for the realization of  efficient single-photon sources by enhancing the emission of light~\cite{Aharonovich2011,Lodahl2015}. 
Such a control of the emission via the modification of the electromagnetic environment is a concept that dates back to the pioneering work of Purcell~\cite{Purcell_1946} performed in the microwave range followed by the experiments of Drexhage \cite{drexhage1970} that demonstrated the possibility of controlling the lifetime of fluorescent molecules in the visible range. This phenomenon is ubiquitous, and it has also been used to control the resonant scattering by dielectric nanorod antennas \cite{holsteen2017}. 

The figure of merit that quantifies the emission enhancement is called the {\it Purcell factor}~\cite{Purcell_1946}, and it is proportional to the $Q$-factor.  Optical nanoantennas were first realized with plasmonic materials \cite{novotny2011,novotny2012}, but recently dielectric resonators have been shown to allow for large enhancements of the Purcell factor via the excitation of both electric and magnetic optically-induced Mie-type resonances~\cite{Albella_2013,zambrana2015}. The excitation of magnetic resonances presents the advantage of enhancing light emission also via the magnetic dipole transitions. This effect was first theoretically predicted  \cite{rolly2012,schmidt2012,zambrana2015} and confirmed later in experiments~\cite{sanz2018,vaskin2019,sugimoto2021}. This is a very promising application for dielectric nanoantennas as the enhancement of light emission empowered by the magnetic dipole resonances is an emerging area of research \cite{karaveli2011,baranov2017}. 
The enhancement of the Purcell factor was used successfully to improve the emission of quantum dots in silicon nanoantennas~\cite{rutckaia2017} and also for metallic and hybrid nanoantennas~\cite{barreda2021}. 
Control of the emission can also be achieved dynamically~\cite{casabone2021}. 
Finally, nanoantennas can also be designed to enhance the performance of quantum emitters, 
providing promising platforms for the realization of single-photon sources~\cite{zalogina2018}.

Bound states in the continuum (BICs) appear as a special type of nonradiating modes associated with an infinite $Q$-factor~\cite{hsu2016}.  Such states can originate from different physical mechanisms~\cite{koshelev2019,tonkaev2020}.  Symmetry protected BICs occur in photonic crystal slabs, and they result from the  impossibility of these modes to couple to propagating fields outside the photonic crystal because of symmetry restrictions~\cite{hsu2016,lee2012}.  Further, the so-called accidental BICs appear from interferences between several resonances~\cite{hsu2016,koshelev2019}. They are observed when a system parameter is varied continuously. This concept was introduced in quantum mechanics where the coupling between resonances is controlled by engineering the potential~\cite{Friedrich1985}.

In optics, one of the first attempts to study BICs was made in the physics 
of photonic crystals~\cite{hsu2013}.  While BICs can be realized in gratings or photonic crystals which are infinite in two directions, it is much more challenging to observe such BICs in compact  structures and even more in subwavelength systems~\cite{hsu2016}.  The existence of BICs, also called {\it embedded eigenstates}, was predicted theoretically in a coated nanosphere where the permittivity of the outer shell vanishes~\cite{monticone2014}. 
In more realistic configurations, it is still possible to take advantage 
of the coupling of resonances in nanostructures to increase the $Q$-factor  of one resonance, even if it does not lead to accidental BICs with infinite $Q$-factors.  In photonics, such an approach was suggested to enhance the $Q$-factors of the modes of optical  micro-cavities~\cite{wiersig2006} and coupled  dielectric  nanopillars~\cite{song2010}.  It was shown recently that high-refractive index nanodisks supporting multiple  resonances are a good platform to employ this approach~\cite{rybin2017,bogdanov2019,koshelev2020c}. 
Due to similarity of this approach with accidental BICs~\cite{Friedrich1985}, the large $Q$-factors achieved through the interference of several resonances are called {\it quasi-BICs}.  Quasi-BICs have been observed experimentally in AlGaAs nanodisks~\cite{melik2021}, and they have been used in various applications~\cite{koshelev2019,koshelev2020b}  including nonlinear optics \cite{carletti2018,carletti2019,koshelev2020} and lasing from a single nanoparticle~\cite{mylnikov2020}.
Compared to photonic crystal cavities, ring resonators, and other setups~\cite{novotny2012}, such compact nanostructures
supporting quasi-BICs exhibit lower Q-factors and Purcell enhancements~\cite{lalanne2018}. However, their relatively small
device footprint allows these resonators to be used, e.g.,
as meta-atoms in metasurfaces~\cite{koshelev2018}.

Unlike BICs, which lead to a perfect confinement of light, quasi-BICs suffer from residual radiation losses.  As a consequence, for a rigorous treatment of quasi-BICs it is important to use quasinormal modes (QNMs) and associated complex eigenfrequencies which generalizes modal approaches to dissipative and non-Hermitian systems~\cite{lalanne2018,kristensen2020,wu2021}. 
The influence of quasi-BICs on light-matter interactions and, in particular, their coupling to a light source can be quantified by using QNM expansions. The QNM analysis of the coupling of an electromagnetic dipole source 
to an optical resonator, i.e., the modal expansion of the Purcell factor, has been carried out through several approaches~\cite{sauvan2013,ge2014,zambrana2015,muljarov2016,Zschiedrich2018}.

In this paper, we study quasi-BICs numerically.  We consider dielectric nanoresonators either {\it with} or {\it without} a substrate and demonstrate that they can support interfering resonances with a strong coupling between a pair of modes.
We choose to design the structure with the refractive index of GaAs. The motivations behind this choice come from the fact that including a dipole emitter into such a structure can be realized using modern nanofabrication methods allowing to include a quantum dot in a GaAs nanodisk~\cite{Kaganskiy_2018}.
We carry out numerical simulations with a localized source embedded into the resonator to demonstrate different physical regimes. Modal expansions of the Purcell factor and far-field patterns reveal a complex interplay between different modal contributions interfering destructively in the spectral vicinity of quasi-BICs, yielding a strong enhancement of the Purcell factor and single  modal excitation when the parameters of the source and resonator are tuned to match the quasi-BIC conditions. 

\begin{center}
\begin{figure}
\includegraphics[width=0.5\textwidth]{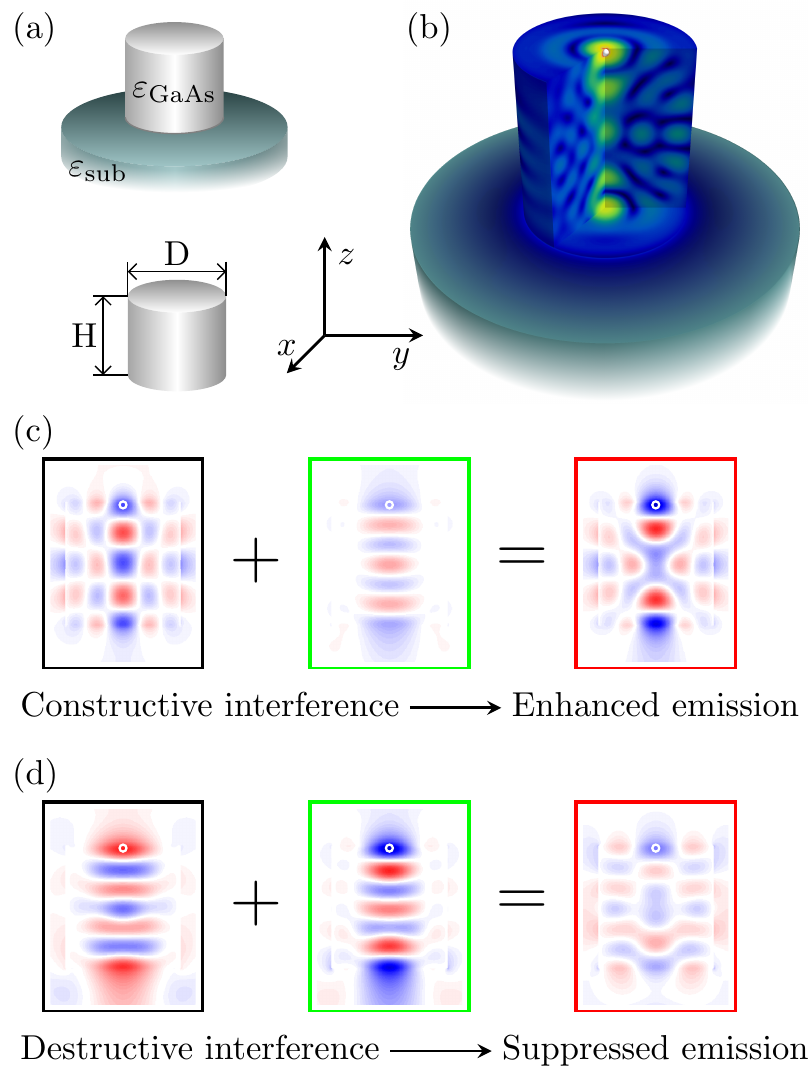}
\caption{Principle of the enhanced and suppressed emission with interfering resonances.  (a) Schematics of GaAs nanodisks with and without a substrate. The aspect ratio $D/H$ is tuned to control the interference between the two main modes of the nanodisk.  (b) Visualization of the electromagnetic field distribution resulting from a  dipole emitter, represented by a white sphere, which is located below the top face of the nanodisk. Its frequency is chosen to excite the two modes of interest.
(c) and (d): 2D cross-sections through the dominant two modal fields (left) and the total field distribution (right) visualizing the real part of the y field component. Red and blue colors correspond to negative and positive fields, respectively.  
The emitter position is indicated with a white circle. 
(c) When the two modal fields are excited in phase they interfere constructively, leading to enhancement of dipole emission. 
(d) Out-of-phase excitation of the two modal fields at a different dipole emission frequency, results in suppressed emission. 
}
\label{Fig:1} 
\end{figure}
\end{center}

The major steps followed in this article are illustrated in Fig.~\ref{Fig:1}. In Sec.~\ref{Sec_quasi_Bic}, we vary the aspect ratio $D/H$ of a GaAs nanodisk to control the interference between the two modes of the nanodisk with or without a substrate. In particular, the strong coupling between these modes leads to the appearance of a high-$Q$ mode: the quasi-BIC resonance. Sec.~\ref{coupling_dipole} considers the coupling of a dipole source with the nanodisk, leading to a complex electromagnetic response as seen in Fig.~\ref{Fig:1}(b). Modal expansions are employed to analyze the role of the interference between the nanodisk modes for the coupling with the dipole. These expansions enable to identify how the constructive interference between two modal contributions leads to the enhancement of the dipole emission, as illustrated in Fig.~\ref{Fig:1}(c). On the other hand, destructive interference leads to the inhibition of the dipole emission, as illustrated in Fig.~\ref{Fig:1}(d). The modal analysis of the radiation pattern is carried out in Sec.~\ref{sec:radiation}. Finally, Sec.~\ref{sec:conclusion} concludes the paper.  

\section{Quasi-BICs in isolated nanodisks}\label{Sec_quasi_Bic}
To understand the appearance of quasi-BIC states, first we review the theoretical approach employed to study the mode  coupling~\cite{wiersig2006,yi2019,heiss2000}.  A good insight in the physics of strong coupling for interfering resonances can be gained 
from a phenomenological model of mode coupling that involves the two modes with the uncoupled eigenfrequencies $\omega_{\text{un},1}$ and $\omega_{\text{un},2}$. When these two eigenfrequencies are far apart in the complex plane, there is no  coupling between them. However, when the eigenfrequencies get close to each other, the coupling has to be taken into account and modifies the trajectories of these eigenfrequencies when a parameter is varied. The eigenfrequencies of the coupled modes can be found as the eigenvalues of an effective two-mode Hamiltonian, and they are equal to
\[
\omega_{\pm} = \left(\frac{\omega_{\textrm{un},1} +
\omega_{\textrm{un},2}}{2}\right)\pm\sqrt{\gamma}, 
\]
where 
\[
\gamma = \left(\frac{\omega_{\textrm{un},1} -
\omega_{\textrm{un},2}}{2}\right)^2 + v^2 \]
with $v$ being the coupling coefficient between the modes~\cite{yi2019}.
We are interested in the regime where these two resonances are close to each other, and therefore we assume that $\Re\left(\omega_{\textrm{un},1}\right) =
\Re\left(\omega_{\textrm{un},2}\right)$ and $v$ is real as in Ref.~\cite{yi2019}. 

\begin{center}
\begin{figure}
\includegraphics[width=0.49\textwidth]{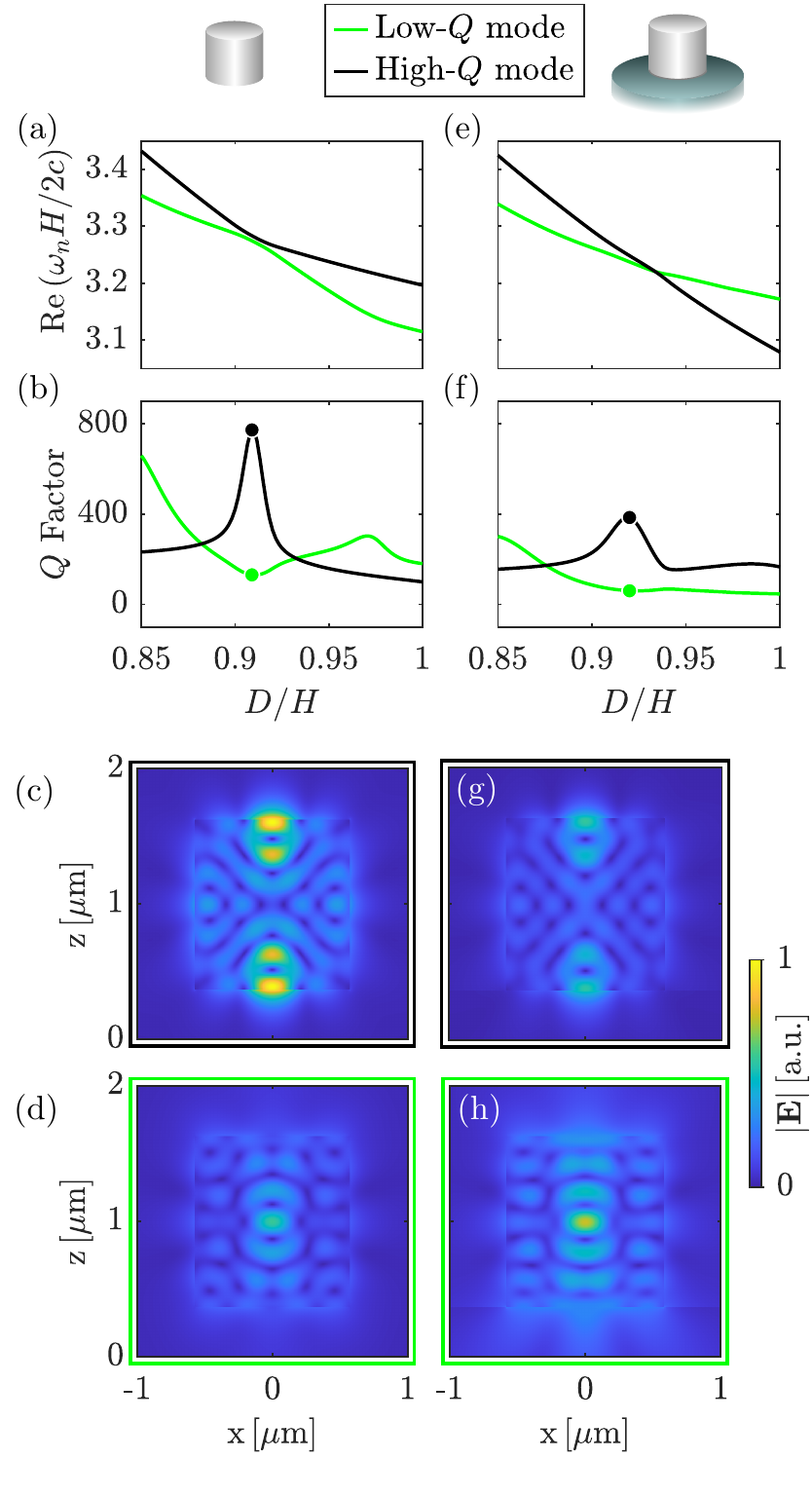}
\caption{
Real parts of two eigenfrequencies of interest (a, e) and corresponding $Q$-factors (b, f) as function of nanodisk aspect ratio $D/H$.
Avoided crossing of the real parts
and local maximum and minimum of the $Q$-factors at $D/H= 0.909$ 
indicate strong coupling for the nanodisk {\it without} substrate (a, b).
Crossing of the real parts of two eigenfrequencies at $D/H= 0.93$ 
and avoided crossing of the $Q$-factor curves at $D/H = 0.944$ leading to a peak at $D/H= 0.92$
indicate weak coupling for the nanodisk {\it with} substrate (e, f).
Field intensity maps $\lvert\mathbf{E}\rvert$ of the QNMs
in an $x-z$-cross section through the 3D field distribution.
(c), resp.~(d), corresponds to the high-$Q$ (resp.~low-$Q$) mode of the isolated nanodisk [at the aspect ratio indicated by the black, resp. green, dot in (b)]. 
(g), resp.~(h), corresponds to the high-$Q$ (resp.~low-$Q$) mode of the nanodisk with substrate [black, resp.~green, dot in (f)].
}
\label{Fig:2} 
\end{figure}
\end{center}

As explained in \cite{wiersig2006,yi2019}, two regimes of the mode coupling may be realized depending on the relation between $v$ and $\frac{1}{2}(\omega_{\mathrm{un},1} - \omega_{\mathrm{un},2})$. 
When  $2v<|\Im\left(\omega_{\textrm{un},1} - \omega_{\textrm{un},2}\right)|$, 
the mode eigenvalues become
\[
\omega_{\pm} = \left(\frac{\omega_{\textrm{un},1} + 
\omega_{\textrm{un},2}}{2}\right)\pm i\sqrt{|\gamma|},
\]
and one observes that the coupling mostly alters the imaginary part of the eigenvalues resulting in an avoided crossing of the imaginary parts of the coupled eigenvalues and a crossing of their real parts.  This behavior is a direct signature of the mode weak coupling. On the other hand, if 
$2v>|\Im\left(\omega_{\textrm{un},1} - \omega_{\textrm{un},2}\right)|$, 
the mode eigenvalues are presented as 
\[
\omega_{\pm} = \left(\frac{\omega_{\textrm{un},1} + 
\omega_{\textrm{un},2}}{2}\right)\pm \sqrt{|\gamma|}, 
\]
suggesting that the coupling of the eigenmodes mostly alters 
the real part of the eigenfrequencies yielding, this time, an avoided crossing of the real parts of the coupled eigenvalues and a crossing of the imaginary parts.
A more detailed discussion on the coupling
regimes between modes for a purely real or a purely imaginary coupling constant can be
found in the~\ref{appendix_sec_coupling}.
In the following, we discuss how the mode coupling may result in the appearance of a hybridized quasi-BIC mode. 

We consider a Gallium Arsenide (GaAs) nanodisk resonator with a height 
$H = 1260\,\mathrm{nm}$ and varying diameter $D$ in two different configurations:
The nanodisk is just surrounded by air (case 1), and, the nanodisk is placed on a glass substrate and surrounded by a super-space of air (case 2). 
The optical properties of the system are investigated in the 
near-infrared wavelength range; the corresponding constant relative permittivities in our model are 
$\epsilon_\textrm{GaAs} = 11.56$, $\epsilon_\textrm{sub} = 2.25$, and
$\epsilon_\textrm{air} = 1.0$.
The time-harmonic optical fields are modeled using Maxwell's equations, 
\begin{equation}
\nabla\times\mu_0^{-1}\nabla\times\mathbf{E}(\mathbf{r},\omega)-
\epsilon(\mathbf{r})\omega^2\mathbf{E}(\mathbf{r},\omega) = 
i\omega\mathbf{J}\left(\mathbf{r}\right),
\label{maxwell}
\end{equation}
where $\mu_0$ is the vacuum permeability, $\epsilon(\mathbf{r})$ is the permittivity, 
and $\mathbf{J}\left(\mathbf{r}\right)$ the source current density.
For numerically solving Eq.~\eqref{maxwell}, we use an adaptive, higher-order finite element method (FEM) \cite{pomplun2007}.
For computing the eigenmodes $\mathbf{E}_n$ of the system and their associated eigenfrequencies $\omega_n$, i.e., solutions to Eq.~\eqref{maxwell}  where $\mathbf{J}=0$, 
the cylindrical symmetry of the system is taken into account.
Only modes with an azimuthal quantum number equal to 1 or -1
are investigated because these are the only ones excited by 
a dipole located on the axis of rotation, which is the configuration 
we are investigating in the second part of this study.
Furthermore, only the component of the polarization normal to the symmetry axis can couple to the modes of interest and therefore we restrict to a polarization with $z=0$. Without loss of generality we chose a y-polarized dipole.

In order to find a quasi-BIC condition, the interference between two modes of the structure has to be tuned \cite{wiersig2006,rybin2017,huang2021}.
This is done by varying the geometry parameter, $D$, and computing eigenmodes $\mathbf{E}_n$ and their associated eigenfrequencies $\omega_n$, where $n$ is the mode index.
Note that alternatively, a perturbation approach based on QNMs may be employed for finding the quasi-BICs \cite{yan2020}. 
Figure~\ref{Fig:2}(a,b,e,f) shows how
the normalized frequency, $\Re(\omega_{n} H/2c)$, and the $Q$-factor, 
\[
Q= -\frac{1}{2} \frac{\Re\left(\omega_n\right)}{\Im (\omega_n)}, 
\]
depend on the aspect ratio $D/H$.
In Fig.~\ref{Fig:2}(a,b), the case where the GaAs nanodisk is located in air is considered. It can be observed that the real part of the eigenfrequencies is showing a repulsion behavior at $D/H= 0.909$ and an almost coinciding peak reaching $Q \approx 800$ is observed for the $Q$-factor of one of the modes while a minimum is seen for the other mode. As discussed above, this behavior is an indication of strong coupling between the two modes. The high-$Q$ mode can thus be considered to be a quasi-BIC. Figures~\ref{Fig:2}(e,f) show the results for the second case, where the nanodisk is put on a glass substrate. It can be observed that, for the investigated modes and parameter range, 
the real part of the eigenfrequencies shows a crossing at $D/H = 0.933$. 
We observe a peak of the $Q$-factor reaching $Q\approx 400$ at $D/H = 0.92$. 
In fact, this peak is linked to the anti-crossing or level-repulsion occurring for the  imaginary parts of the eigenfrequencies. This avoided crossing of the imaginary parts of the eigenfrequencies shows up in  Fig.~\ref{Fig:2}(f) at about $D/H=0.944$. The qualitative analysis based on the effective Hamiltonian discussed above  shows that this behavior is an indication of weak coupling between the two modes. The transition from strong to weak coupling when a substrate is added indicates that there must be an exceptional point, i.e., a condition for which the two coupled eigenvalues would become degenerated \cite{heiss2012}, when continuously varying the refractive index of the substrate from 
1 to 1.5 \cite{yi2019,heiss2000,rodriguez2016,deng2021}. 
To conclude the discussion on the avoided crossing of the eigenfrequencies, 
we show, in Figs.~\ref{Fig:2}(c,d,g,h), the field patterns associated with both modes when the $Q$-factor is maximized. For the case without substrate, this occurs for $D/H= 0.909$ while, when the substrate is added, the maximum occurs for an aspect ratio of $D/H=0.92$. This helps to understand the level repulsion observed since, in both cases, the modes have apparently very different field patterns: The high-$Q$ mode field is concentrated in hot spots located at the top and bottom of the disk while, for the low-$Q$ mode, it is concentrated at the center of the disk. This apparent difference in the localization of the modes certainly prevents their merging. 

\begin{figure*}
\includegraphics[width=0.9\textwidth]{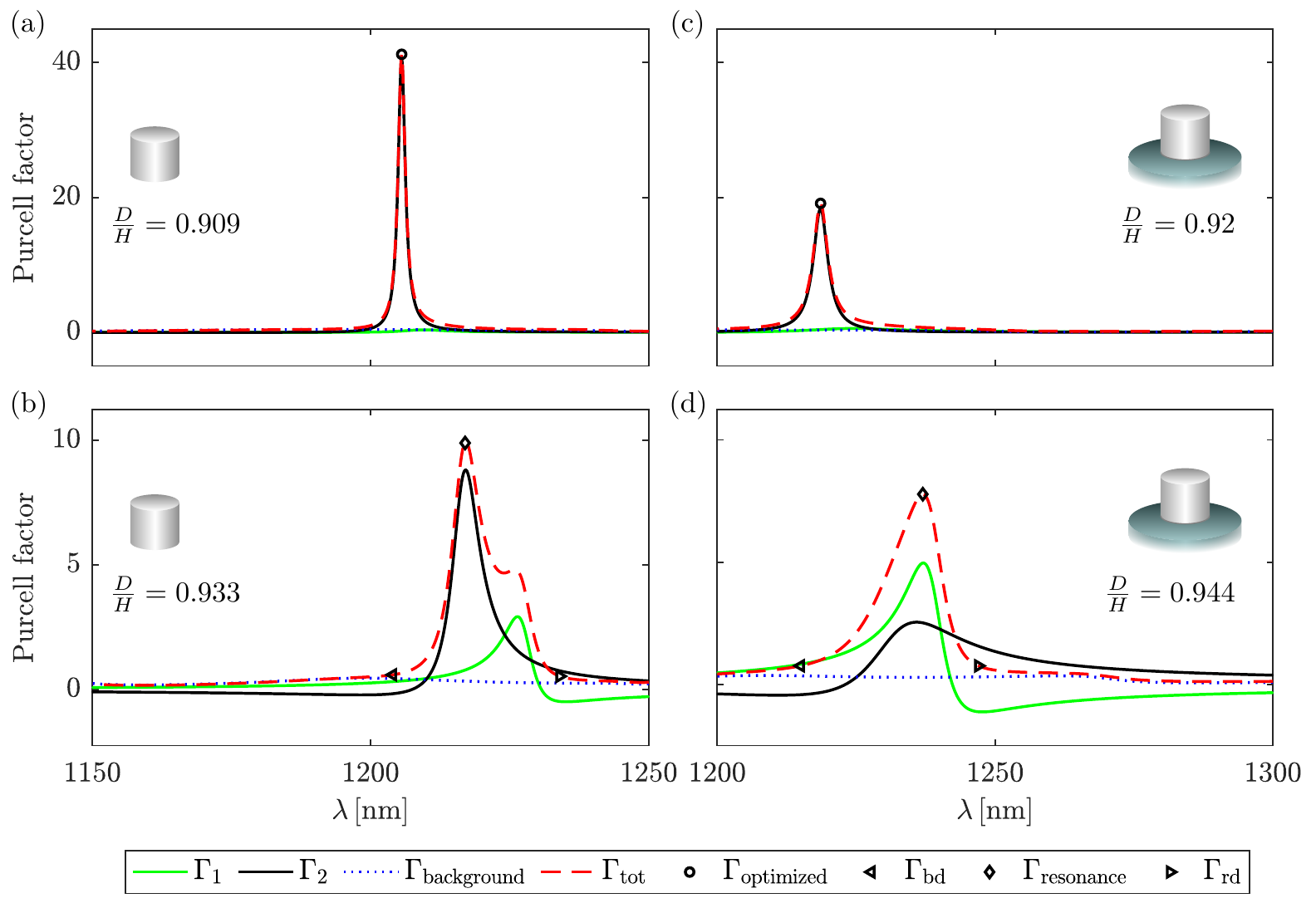}
\caption{Modal analysis of the wavelength ($\lambda$) dependent Purcell factor $\Gamma$ for a $y$-polarized dipole located on the symmetry axis $20\,\mathrm{nm}$ and $27\,\mathrm{nm}$ below the top face of the nanodisk in the case without substrate (a,b) and with substrate (c,d), respectively.  (a)~Modal expansion for the aspect ratio $D/H= 0.909$ (maximum $Q$-factor in Fig.~\ref{Fig:2}(b). The high-$Q$ mode corresponds to the modal Purcell factor $\Gamma_2$ (a black solid curve) and is solely responsible for the peak of the total Purcell factor $\Gamma_\mathrm{tot}$ (dashed red curve) at around $1205\,\mathrm{nm}$.  The contributions of the low-$Q$ mode $\Gamma_1$ (green solid curve) and of the background $\Gamma_{\text{background}}$ (dotted blue line) are negligible.
(b)~Modal expansion for $D/H= 0.933$  (crossing of $Q$-factors in Fig.~\ref{Fig:2}(b).  Both modal terms $\Gamma_1$ and $\Gamma_2$ are of the same order of magnitude and destructively interfere in regions where they are of different sign.  The impact of $\Gamma_{\text{background}}$ is constant and negligible in resonant regions.  (c)~Modal expansion for $D/H= 0.92$  (peak of the high $Q$-factor in Fig.~\ref{Fig:2}(f). The high-$Q$ mode corresponds to $\Gamma_2$ and is responsible for the peak of $\Gamma_\mathrm{tot}$ at around $1220\,\mathrm{nm}$.  (d)~Modal expansion for $D/H= 0.944$ (avoided crossing of $Q$-factors in Fig.~\ref{Fig:2}(f). The modal terms interfere, as in (b).
The markers $\Gamma_{\text{optimized}}$, $\Gamma_{\text{bd}}$, $\Gamma_{\text{resonance}}$ and $\Gamma_{\text{rd}}$ indicate the wavelengths for which far-field patterns are displayed in Fig.~\ref{Fig:4}.}
\label{Fig:3} 
\end{figure*}

\section{Coupling of a point source to a nanoresonator}
\label{coupling_dipole}

Now, we turn to the study of a dipole emitter coupled to the investigated 
nanoresonator considering the two cases, the nanoresonator with and without substrate.  It is worth noting that the coupling of a dipole with a BIC in an array of  nanoparticles have already been studied \cite{abujetas2021},
but we will here focus on the coupling of a dipole with the quasi-BIC 
arising in an individual nanodisk. We consider Maxwell's equations, given by Eq.~\eqref{maxwell}, with the 
current  density $\mathbf{J} = \mathbf{j}\delta\left(\mathbf{r}-\mathbf{r}_\mathrm{d}\right)$ that is a point source located at 
$\mathbf{r}_\mathrm{d}$. 
The Purcell factor, which is used to quantify the enhancement of the emission, is defined as
$\Gamma(\omega) = 
-[\Re\left(\mathbf{E}(\omega,\mathbf{r}_\mathrm{d})\cdot
\mathbf{j}^{*}(\omega,\mathbf{r}_\mathrm{d})\right)]/[2\Gamma_\mathrm{b}(\omega)]$, 
where $\Gamma_\mathrm{b}(\omega)$ describes the emission of the dipole in a homogeneous medium of the permittivity $\epsilon_\textrm{GaAs}$. The interest of studying the Purcell factor and its modal analysis is twofold. On the one hand, one can see how a mode with a $Q$-factor as large as the one of the quasi-BIC can affect the dipole emission. On the other hand, looking at the modal analysis of the Purcell factor would allow to use it as a probe to study the interplay between several modes. This is particularly interesting for quasi-BICs since interferences between modes are at the origin of their formation. 

To do so, we start by considering the Purcell factor for a dipole located at the  maximum of the field amplitude of the high-$Q$ mode. This position is on the symmetry axis of the nanodisk, about $30\,\mathrm{nm}$ below the top face.

The consequences of the interplay between resonances at the origin of the quasi-BIC can  be better understood by carrying out a modal analysis of the Purcell factor. 
Our method for deriving modal expansions relies on the use of 
Riesz projections \cite{Zschiedrich2018,Binkowski2020}. 
The modal expansion of the Purcell factor reads as 
\begin{equation}
\Gamma_\textrm{tot}(\omega) = \sum_{n=1}^2 \Gamma_n(\omega)+\Gamma_\textrm{background}(\omega),
\end{equation}
where $\Gamma_n$ are the modal contributions to the Purcell factor that are 
computed using contour integrals around the eigenfrequencies.  Here, we take into account only the two interfering modes, i.e., the modes 
which are also shown in Fig.~\ref{Fig:2}. The modal Purcell factors $\Gamma_{1}$ and $\Gamma_{2}$ are contributions corresponding to  these two modes.  The term $\Gamma_\textrm{background}$ contains the contributions of all other poles as well as the nonresonant background \cite{Zschiedrich2018,Binkowski2020}. Finally, $\Gamma_{\textrm{tot}}$ corresponds to the total expansion including both the modal and background contributions. The different black markers indicate the wavelengths at which the radiation patterns are computed in Fig.~\ref{Fig:4}. Details about the modal expansions are provided in the~\ref{appendix_sec_rp}.

First, we look at the coupling of the dipole to the nanoresonator with the 
geometry corresponding to the maximum of the $Q$-factor in Fig.~\ref{Fig:2}.
The results of the modal analysis of the Purcell factor are displayed in Fig.~\ref{Fig:3}(a), for a nanodisk in air with an aspect ratio 
$D/H = 0.909$, and, in Fig.~\ref{Fig:3}(c), for a nanodisk on a substrate 
with an aspect ratio $D/H = 0.92$. In both cases, the peak observed in the Purcell factor spectrum can be directly linked to the modal contribution corresponding to the high-$Q$ mode.  In the region around the peak, the contributions from the low-$Q$ mode and the background are very small or even negligible. This demonstrates that, for a resonator supporting a quasi-BIC, an emitter may easily excite nearly exclusively this resonance. 
We note that the quasi-BIC allows to reach a high Purcell factor of $\Gamma \approx 40$ in the case without substrate and $\Gamma \approx 20$ in the case with substrate. 

It is also worth looking at configurations where one can expect that the contributions to the Purcell factor from the two main modes would be of the same order of magnitude. This would allow us to investigate the interplay between modal contributions. This is the reason for showing, in Fig.~\ref{Fig:3}(b), the Purcell factor for an aspect ratio of $D/H = 0.933$ for the disk without substrate corresponding to the crossing of the $Q$-factor of the two modes in Fig.~\ref{Fig:2}(b). For the case with substrate, we consider the aspect ratio $D/H = 0.944$ as it corresponds to the avoided crossing of the imaginary parts of the eigenvalues as can be seen from the $Q$-factor trajectories in Fig.~\ref{Fig:2}(f). This avoided crossing is caused by the interference of the interacting modes. Therefore, we expect that the interference will be seen in the contributions of the modal expansion. 
The Purcell factor again shows a distinct maximum, with a value of $\Gamma \approx 8$ with substrate and $\Gamma \approx 10$ without substrate. 
However, as expected, both modal contributions have the same order of magnitude. Also, the qualitative shape of both spectra of the modal Purcell factors are approximately mirror-symmetric to each other with respect to the resonance wavelength.  This behavior yields the fact that, away from the resonance, the signs of the modal  contributions of the two modes are opposite, leading to destructive interference  in these spectral regions. This is the case in Fig.~\ref{Fig:3}(b) for wavelengths below $\sim 1210\,\mathrm{nm}$ and above $\sim 1230\,\mathrm{nm}$. For the case including a substrate in Fig.~\ref{Fig:3}(d), we observe a similar behavior for wavelengths below $\sim 1225\,\mathrm{nm}$ and above $\sim 1245\,\mathrm{nm}$. The destructive interference between modes has been used previously to qualitatively  describe the appearance of quasi-BICs \cite{koshelev2020b}. In the present study, we show that the effect can be quantified by using modal expansion techniques. 

Note that the interplay between the modes at the optimal aspect ratio becomes visible when the position of the dipole is moved away from the hot spot of the high-Q mode. Corresponding simulation results can be found in the \ref{appendix_sec_dipole}.

\section{Modal analysis of radiation patterns} 
\label{sec:radiation}

\begin{figure*}
\includegraphics[width=1\textwidth]{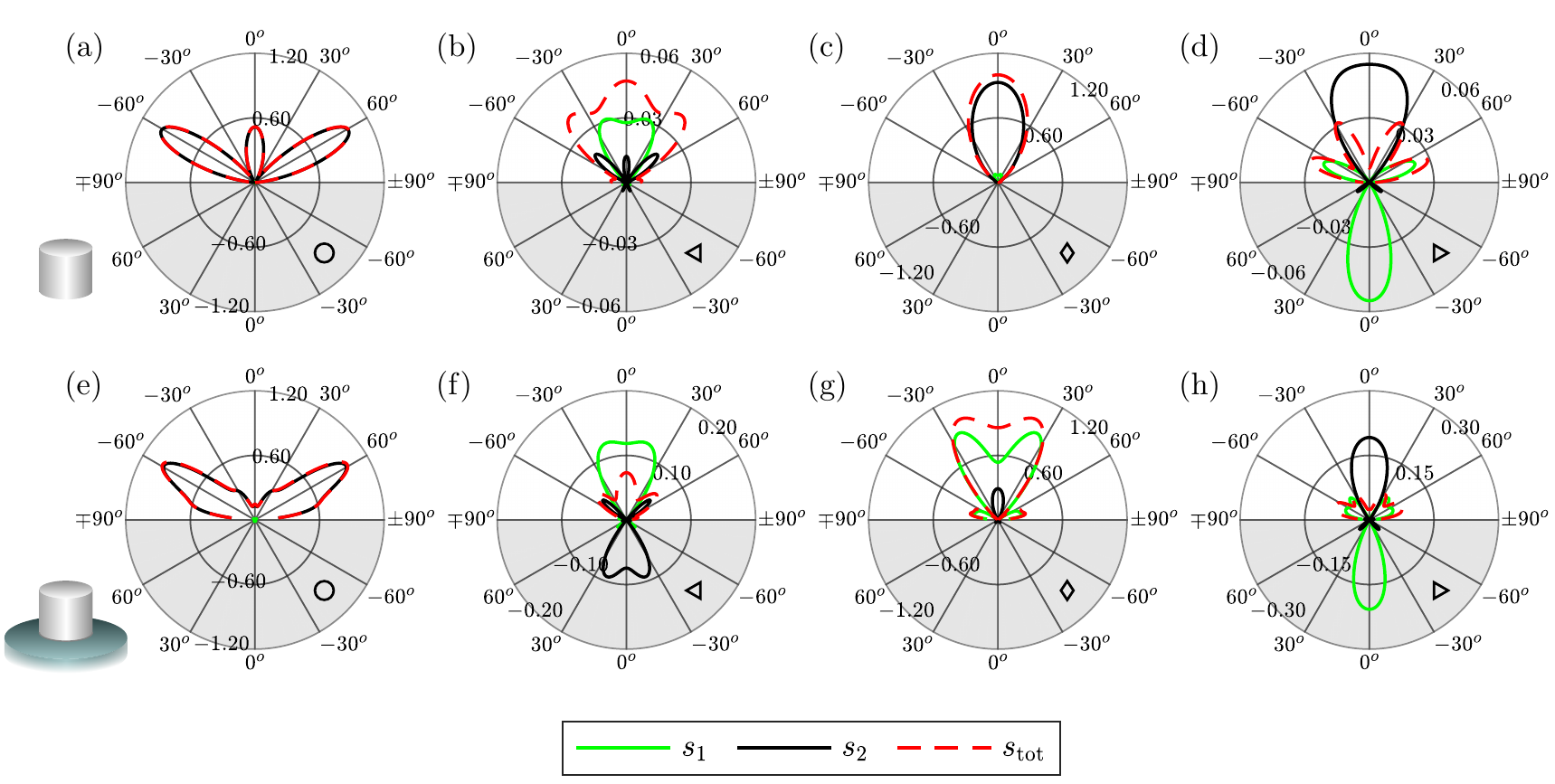} 
\caption{Modal decomposition of the $\theta$-dependent, normalized radiation patterns towards the top 
for a dipole on the symmetry axis $20\,\mathrm{nm}$ and $27\,\mathrm{nm}$ below the top face of the nanodisk in the case without substrate (a-d) and with substrate (e-h), respectively.
The green (black) solid curve corresponds to the angle-resolved, far-field modal energy flux $s_1$ ($s_2$) corresponding to the low-$Q$ (high-$Q$) mode.
The red dashed curve corresponds to the total energy flux $s_{\text{tot}}$.
The upper half of each diagram shows positive contributions while the lower  half in gray shows negative contributions.
The dipole emission wavelengths correspond to the different $\Gamma$ markers in Fig.~\ref{Fig:3} which are reproduced in the right bottom of each emission diagram.  
(a, e) show the on-resonant far field radiation for nanodisks supporting the quasi-BIC 
($D/H= 0.909$ and $\lambda = 1206\,\mathrm{nm}$, resp.~$D/H = 0.92$ and $\lambda = 1219\,\mathrm{nm}$) with clearly dominating contribution from the high-$Q$ mode. 
(b-d), resp.~(f-h) show the far field radiation for nanodisks with aspect ratios of  
$D/H = 0.933$, resp.~$D/H= 0.944$ (i.e., at the avoided crossing, resp.~crossing of the eigenfrequencies, cf., Figs.~\ref{Fig:2}(a, e) for on-resonant sources $(\lambda = 1217\,\mathrm{nm}$/$1237\,\mathrm{nm}$ in c/g) and off-resonant sources ($\lambda = 1204\,\mathrm{nm}/1234\,\mathrm{nm}/1219\,\mathrm{nm}/1247\,\mathrm{nm}$ in b/d/f/h).
While for on-resonant sources a single mode is predominantly contributing to the far field pattern (c, g), in off-resonant settings, the two relevant modes can interfere constructively (b) or destructively (d, f, h), as can be seen from the equal or different signs of the two dominant modal contributions in each case. 
}
\label{Fig:4} 
\end{figure*}

It is well known that the coupling with nanostructures can alter the radiation pattern  of a quantum emitter~\cite{novotny2011}. This ability to control the emission pattern of a dipole emitter with nanostructures  has a great practical interest since it can improve the collection of the emitted  field with an optical system.  A modal analysis allows to understand how each mode but also the interferences between  modes modifies the emission pattern. 
We will consider the far-field pattern of the energy flux density defined as
$s(\mathbf{r},\omega) = 
\frac{1}{2}\Re\left(\mathbf{E}^{*}(\mathbf{r},\omega)\times
\frac{1}{i\omega\mu_0}\nabla\times\mathbf{E}(\mathbf{r},\omega)\right)\cdot\mathbf{n}$, 
i.e., the projection of the Poynting vector on the normal vector $\mathbf{n}$ 
in the direction of field propagation.  The modal expansion of $s(\mathbf{r},\omega)$ is computed using Riesz 
projections~\cite{Binkowski2020,betz2021} leading to the expression
$s(\mathbf{r},\omega)=\sum_{n = 1}^2 s_n(\mathbf{r},\omega)+
s_{\text{background}}(\mathbf{r},\omega)$, where $\mathbf{r}$ is a point located in the far-field. We will in particular look at the dependency of the radiation pattern with $\theta$ in the x-z plane.  In Fig.~\ref{Fig:4}, the field patterns radiated by the dipole upwards towards the air are plotted for different wavelengths and for different aspect ratios. In Figs.~\ref{Fig:4}(a-d), results are shown for the nanodisk without substrate for aspect ratios equal to $0.909$ and $0.933$ while Figs.~\ref{Fig:4}(g-h) display results for the nanodisk on a substrate for aspect ratios equal to $0.92$ and $0.944$. Please note that the lower region of the plot shown in gray does not correspond  to the field radiated downwards but to the negative modal contributions. Negative contributions are particularly important here, since, as for the Purcell factor,  they are linked to the interferences between different modal contributions. Radiation pattern towards the substrate are actually shown in the supplemental material.
In Figs.~\ref{Fig:4}(a,e), we show the radiation pattern and its modal expansion at the aspect ratio and wavelength of the quasi-BIC. 
Just like for the Purcell factor, one mode has a much larger $Q$-factor than the other,  it is not surprising to find that the radiation pattern can then be almost entirely  understood from the contribution of the high-$Q$ mode while the contributions from the low-$Q$ mode is negligible compared to the contribution of the high-$Q$ mode. 
The results of the modal expansion of the radiation pattern for the nanodisk without substrate with the aspect ratio equal to $0.933$ are plotted in Figs.~\ref{Fig:4}(b-d). These computations are made for the wavelengths on both sides of the main peak in Fig.~\ref{Fig:3}(b), with $1204\,\mathrm{nm}$, $1217\,\mathrm{nm}$, and $1234\,\mathrm{nm}$ in Figs.~\ref{Fig:4}(b, c, d), respectively.

For $\lambda = 1204\,\mathrm{nm}$, we obtain a positive contribution for both the modes summing up to a radiation lobe between $\sim \pm 45^\circ$. For $\lambda = 1217\,\mathrm{nm}$, the main contribution is from mode 2 leading to a quite directional emission between $\sim \pm 30^{\circ}$. Eventually, for $\lambda = 1234\,\mathrm{nm}$, an interference between the two main modal contributions is observed with a positive contribution from mode 2 between $\pm 30^{\circ}$ and a negative contribution of mode 1 in the same range.
In Figs.~\ref{Fig:4}(f-h), we show the results of the modal expansions for the nanodisk on substrate with $D/H=0.944$ for the wavelengths $1215\,\mathrm{nm}$, $1237\,\mathrm{nm}$, and $1247\,\mathrm{nm}$, respectively.  In Fig.~\ref{Fig:4}(f), for $\lambda = 1215\,\mathrm{nm}$, the mode 1 has a positive contribution for angles between roughly 30 and -30 degrees while the mode 2 has a negative 
contribution in the same range of angles.
Consequently, the total radiation pattern is suppressed, resulting from the interference between several modes. A very analogous behavior is observed at $\lambda = 1247\,\mathrm{nm}$ in Fig.~\ref{Fig:4}(h),
however, in this case, the mode 1 has a negative contribution while the mode 2  has a positive contribution. There is, again, a strong interference between the two modes and the far-field pattern cannot  be understood without taking this interference into account.  Finally, in Fig.~\ref{Fig:4}(g), for $\lambda = 1237\,\mathrm{nm}$ corresponding to the 
peak of the Purcell factor in Fig.~\ref{Fig:3}(b),  we observe that the contribution from both modes of interest add up leading to a larger amplitude of the radiation by the dipole and to a confined far-field pattern. 

\section{Conclusions}\label{sec:conclusion}
We have numerically analyzed dielectric nanodisk resonators which support multiple resonances in overlapping frequency ranges.
Using a finite-element-method-based framework, regimes where the resonators support quasi-BIC resonances have been investigated. 
The impact of the resonances on the Purcell factor describing the emission enhancement of a localized source has been shown in 
the quasi-BIC regime as well as in adjacent parameter regimes where several competing resonances are excited. 
The modal contributions to the Purcell factor have been computed using the Riesz projection method, and it has been shown that 
a single QNM causes the strongly enhanced dipole emission in the quasi-BIC situation. 
Further, we have investigated the modal, angular resolved far-field spectrum in on-resonance as well as off-resonance conditions. 
This demonstrated that modal interference strongly impacts both, far-field emission strength as well as angular resolved radiation patterns. 
It has been shown that micron-scale dielectric resonators supporting quasi-BICs allow for 
high Purcell enhancement as well as for highly directed emission of light. 
We expect that, apart from the gained insight in the complex interference behavior in multi-modal resonators, these findings 
will allow for the design of efficient and robust future photonic components, 
such as single-photon emitters for quantum technology applications.\\
\\
\noindent {\it Research Data:} The source code and data for performing the numerical
simulations and producing the resulting figures as reported in
this article will be made available~\cite{colom_binkowski_betz_data}.

\section*{Acknowledgments}
The authors acknowledge funding from the German Research Foundation 
(DFG, Excellence Cluster MATH+, EXC-2046/1, project 390685689), 
the Helmholtz Association (Helmholtz Excellence Network SOLARMATH, project ExNet-0042-Phase-2-3), 
and the German Federal Ministry of Education and Research  (BMBF Forschungscampus  MODAL, project 05M20ZBM). 
Also, this project has received funding from the EMPIR program( European Metrology Programme for Innovation and Research) co-financed  
by the Participating States and by the European Union’s Horizon 2020  
research and innovation program (projects 20FUN05 SEQUME \& 20FUN02 POLIGHT). 
Y.K.~acknowledges support from the Australian Research Council (grants DP200101168 and DP210101292).



\onecolumngrid

\begin{center}
\section*{Appendix}
\end{center}

\setcounter{section}{0}
\setcounter{equation}{0}
\renewcommand\thesection{Appendix \Alph{section}}
\renewcommand\thesubsection{\Alph{section}.\arabic{subsection}}
\renewcommand\theequation{S\arabic{equation}}


\section{Riesz projection principle and results}\label{appendix_sec_rp}

Our approach for carrying out modal expansions relies on Riesz projections~\cite{Zschiedrich2018,Binkowski2020,betz2021}. In a first step, the quantity of interest at real frequencies $\omega_0$ is expressed as a contour integral using Cauchy's integral formula. To this end, it has to be analytically continued to the complex frequency plane. In a second step, the resonance expansion is obtained by deforming the contour around $\omega_0$ until it encloses neighboring poles of the physical system and by the application of Cauchy's residue theorem. Each summand of the expansion corresponds to a contour integral. 

\begin{center}
\begin{figure}[b]
\includegraphics[width=0.9\textwidth]{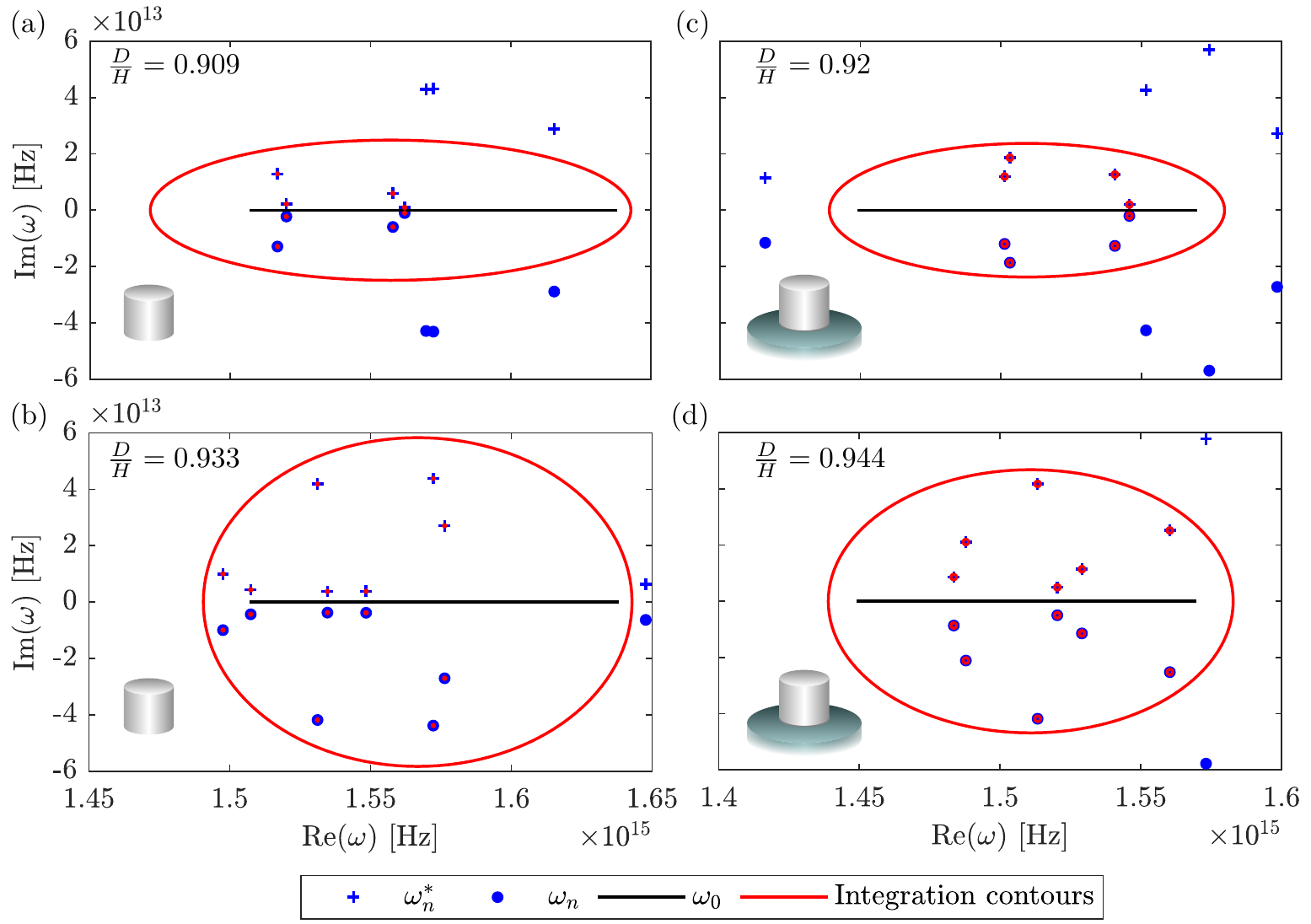}
\caption{Contours used for the modal expansions of Purcell factor and emission pattern of a nanodisk with  different aspect ratios $D/H$, either isolated or placed on a substrate. While the emission pattern is based on a quadratic form and requires contours around the complex conjugate resonance frequencies $\omega_n^*$, which are the poles of $\mathbf{E}^\circ(\omega)$, the circular contours in the upper half space can be ignored for the Purcell factor.}
\label{Fig:SI1} 
\end{figure}
\end{center}

Using the example of the Purcell factor $\Gamma(\omega_0) = -\frac{1}{2}\Re\left(\mathbf{E}(\omega_0,\mathbf{r}_d)\cdot\mathbf{j}^{*}(\omega_0,\mathbf{r}_d)\right)/\Gamma_b(\omega_0)$,
whose expansion is shown in Fig.~3 of the main document, 
the first step yields
\begin{equation}
    \Gamma(\omega_0) =-\frac{1}{2\Gamma_b(\omega_0)}\oint_{C_0}\frac{\Re\left(\mathbf{E}(z,\mathbf{r}_d)\cdot\mathbf{j}^{*}(z,\mathbf{r}_d)\right)}{z-\omega_0}dz, \nonumber
\end{equation}
where $C_0$ is a contour around $\omega_0$. The second step results in the desired expansion of the Purcell factor,
\begin{equation}
    \Gamma(\omega_0) = \sum_n\Gamma_n(\omega_0)+\Gamma_{\text{background}}(\omega_0), \nonumber
\end{equation}
with
\begin{align*}
    \Gamma_n(\omega_0) =&-\frac{1}{2\Gamma_b(\omega_0)}\oint_{C_n}\frac{\Re\left(\mathbf{E}(z,\mathbf{r}_d)\cdot\mathbf{j}^{*}(z,\mathbf{r}_d)\right)}{z-\omega_0}dz \quad \text{and}\\
    \Gamma_{\text{background}}(\omega_0) =& -\frac{1}{2\Gamma_b(\omega_0)}\oint_{C_{\text{background}}}\frac{\Re\left(\mathbf{E}(z,\mathbf{r}_d)\cdot\mathbf{j}^{*}(z,\mathbf{r}_d)\right)}{z-\omega_0}dz.
\end{align*}
The contour $C_n$ is the contour around the nth pole and $C_{\text{background}}$ is the large outer contour. Please refer to Fig.~\ref{Fig:SI1} and note that, for quantities linear in the electric field, such as the Purcell factor in the given form, the complex conjugate poles located in the upper half of the complex plane can be ignored. The integrals are computed numerically using the trapezoidal rule for the used
circular and ellipsoidal contours.

\begin{center}
\begin{figure}
\includegraphics[width=\textwidth]{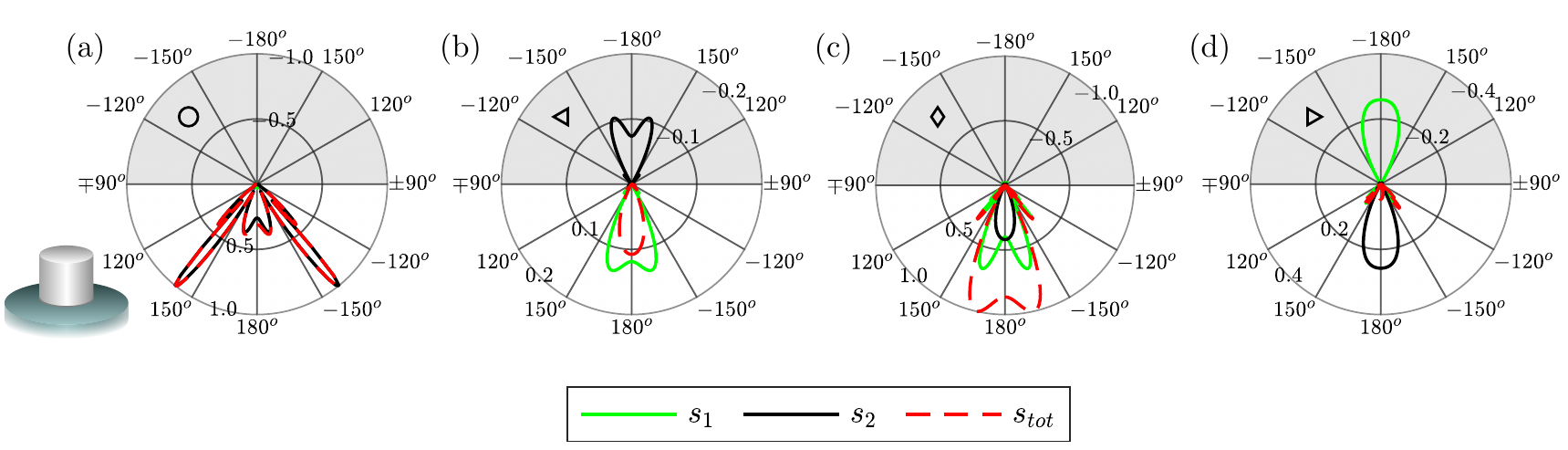}
\caption{Modal analysis of the radiation pattern towards the substrate for a dipole located on the symmetry axis $27\,\mathrm{nm}$ below the upper base of the nanodisk.
  The black markers refer to Fig.~\ref{Fig:3} 
  where they mark the corresponding wavelengths. The optimized system ($D/H=0.92$) shown in (a) illustrates the dominance of a single mode. For (b)-(d) the aspect ratio is $D/H=0.944$. Here, the radiation pattern results from the interference between two dominant modes. In (b), the pattern is shown at a wavelength blue shifted from the maximal Purcell enhancement, in (c), at the maximum and, in (d), at a red shifted wavelength.}
\label{Fig:SI2} 
\end{figure}
\end{center}

In Section~\ref{sec:radiation}, 
we expand the far-field pattern of the radiated flux which is quadratic in the electric field, $s\left(\mathbf{E}(\omega),\mathbf{E}^{*}(\omega)\right) = \frac{1}{2}\Re\left(\mathbf{E}^{*}(\omega)\times\frac{1}{i\omega\mu_0}\nabla\times\mathbf{E}(\omega)\right)\cdot\mathbf{n}$, and hence involves its complex conjugate. The method for expanding quadratic forms was developed in \cite{Binkowski2020}. The application of Cauchy's residue theorem requires a holomorphic expression and therefore does not allow for complex conjugation. As the electric field is a real quantity in the time domain, we have $\mathbf{E}^{*}(\omega)= \mathbf{E}(-\omega)$ for real $\omega$. With the analytic continuation to the complex plane $\mathbf{E}^{\circ}(\omega)$ of $\mathbf{E}(-\omega)$, the holomorphic expression $s\left(\mathbf{E}(\omega),\mathbf{E}^{\circ}(\omega)\right) = \frac{1}{2}\Re\left(\mathbf{E}^{\circ}(\omega)\times\frac{1}{i\omega\mu_0}\nabla\times\mathbf{E}(\omega)\right)\cdot\mathbf{n}$ is defined. The poles of $\mathbf{E}^{\circ}(\omega)$ are located in the upper part of the complex plane. They are the complex conjugates of the resonance poles associated with $\mathbf{E}(\omega)$ as shown in Fig.~\ref{Fig:SI1}. The expansion of $s\left(\mathbf{E}(\omega),\mathbf{E}^{\circ}(\omega)\right)$ consequently features resonant terms from poles in the lower and the upper part of the complex plane. Following this approach~\cite{Binkowski2020}, one can derive the expansion of $s\left(\mathbf{E}(\omega),\mathbf{E}^{\circ}(\omega)\right)$,

\begin{equation}
\begin{aligned}
   s\left(\mathbf{E}(\omega_0),\mathbf{E}^{\circ}(\omega_0)\right) =& -\sum_n\frac{1}{2i\pi}\oint_{C_n}\frac{s\left(\mathbf{E}(z),\mathbf{E}^{\circ}(z)\right)}{z-\omega_0}dz
   -\sum_n\frac{1}{2i\pi}\oint_{C^*_n}\frac{s\left(\mathbf{E}(z),\mathbf{E}^{\circ}(z)\right)}{z-\omega_0}dz \nonumber \\\
   &+\frac{1}{2i\pi}\oint_{C_\mathrm{background}}\frac{s\left(\mathbf{E}(z),\mathbf{E}^{\circ}(z)\right)}{z-\omega_0}dz, \nonumber
\end{aligned}
\end{equation}

where $C_n$ is again the contour around the nth pole and $C_\mathrm{background}$ is the large background contour.
As mentioned above, we have to add the contours around poles in the upper part of the complex plane, which we denote by $C^*_n$.
The expansion of the radiation pattern towards the air is discussed in Section~\ref{sec:radiation}. 
Here, for the sake of completeness, we show the expansion of the radiation towards the substrate at the same wavelengths and aspect ratios as in Fig.~\ref{Fig:4}. 
Overall, the same behavior is observed for the flux radiated towards the substrate as it was for the flux radiated towards the air. In Fig.~\ref{Fig:SI2}(a), one can see that the modal contribution from the high-$Q$ mode dominates all the other contributions. Figures~\ref{Fig:SI2}(b-d) show once more that the radiation pattern of the flux results from the interference of the two main modes. In Figs.~\ref{Fig:SI2}(b) and (d), they interfere destructively and, in Fig.~\ref{Fig:SI2}(c), constructively.


\section{Coupling of resonances}\label{appendix_sec_coupling}

In the main text, we employed a phenomenological method to study the mode coupling. We used the following expressions for the coupled eigenfrequencies:
\begin{equation}
\omega_{\pm} = \left(\frac{\omega_{\textrm{un},1} +
\omega_{\textrm{un},2}}{2}\right)\pm\sqrt{\gamma}, 
\end{equation}
where 
\begin{equation}
\gamma = \left(\frac{\omega_{\textrm{un},1} -
\omega_{\textrm{un},2}}{2}\right)^2 + v^2
\label{def_gamma}
\end{equation}
with $v$ being the coupling coefficient. While we then focused on explaining the crossings and avoided crossings of real and imaginary parts of the resonances as a consequence of different coupling regimes, these formulas can provide further insight into the couplings between resonances. Some additional results based on these formulas are provided in the following.\\

We will study the following uncoupled resonance frequencies: 
\begin{equation}
\begin{aligned}
\omega_{\textrm{un},1} &= 1 - i 0.01\\
\omega_{\textrm{un},2} &= 1 + \Delta - i(0.01 + \Delta\omega_i)
\label{Def_omega_un}
\end{aligned}
\end{equation}
For this study, we keep a fixed value of $\Delta\omega_i = 0.0025$ and study the trajectories of the coupled eigenvalues when $\Delta$ is varied. The impact of the value of $v$ on the coupling of resonances will be studied for two cases: in the first case $v$ is real-valued and positive and in the second case $v^2$ is purely imaginary. We will also study the impact of the coupling of the resonances on their respective Q-factor.

\subsection{Coupling of resonance for $v^2$ real and positive} 

\begin{figure*}
\includegraphics[width=1\textwidth]{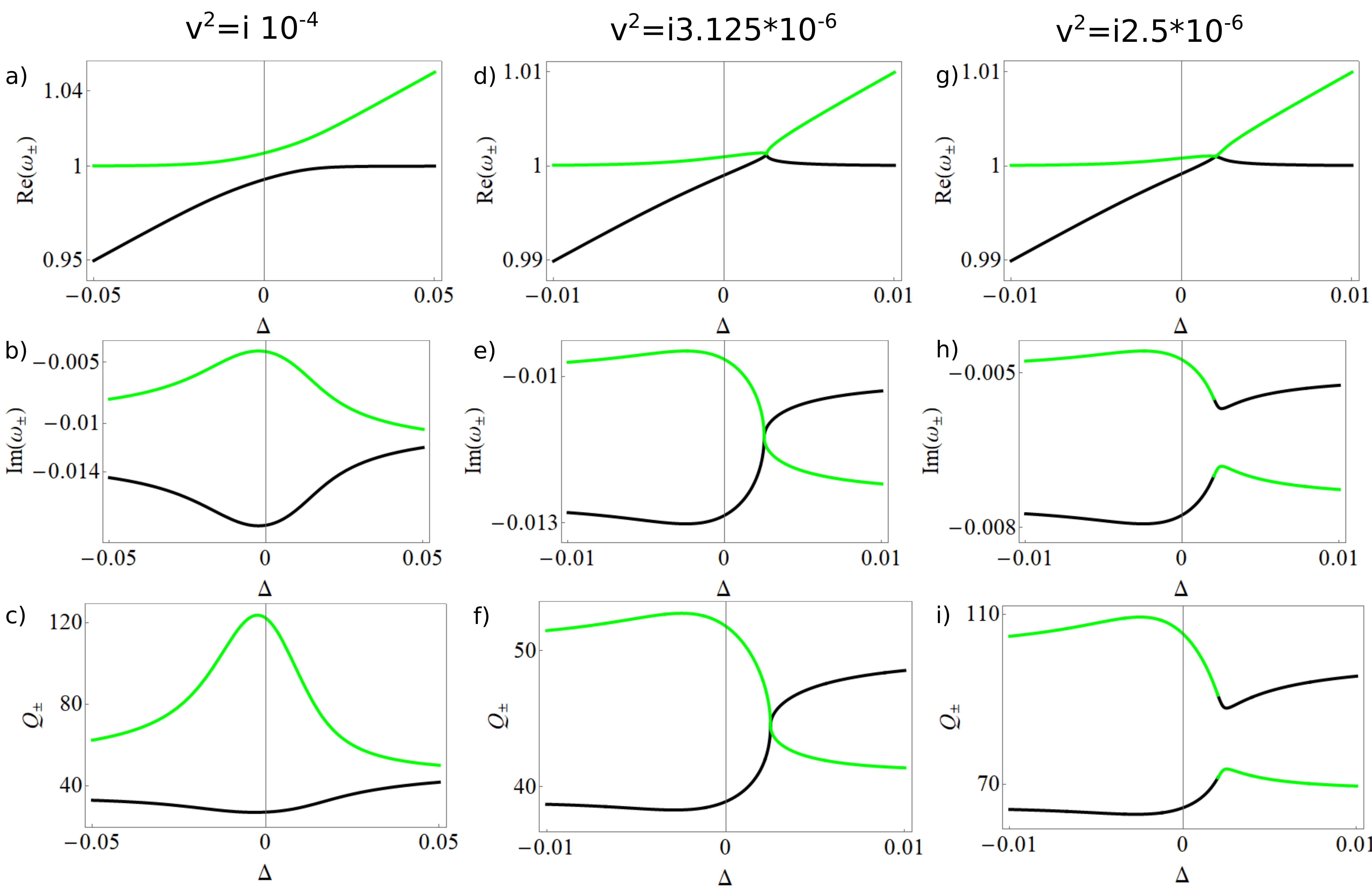} 
\caption{Real parts $\mathrm{Re}(\omega_\pm)$,
  imaginary parts $\mathrm{Im}(\omega_\pm)$ and Q-factors $Q_\pm$ of the two coupled frequencies $\omega_\pm$ as
  a function of the difference of the real parts of the uncoupled eigenfrequencies
  $\Delta = \mathrm{Re}(\omega_{un,2}-\omega_{un,1})$.
  The trajectories are given for three different couplings $v^2$, with $v$ being real and positive.
  For better readability a vertical line is shown at $\Delta = 0$.}
\label{Fig:SI_1} 
\end{figure*}

\begin{figure*}
\includegraphics[width=1\textwidth]{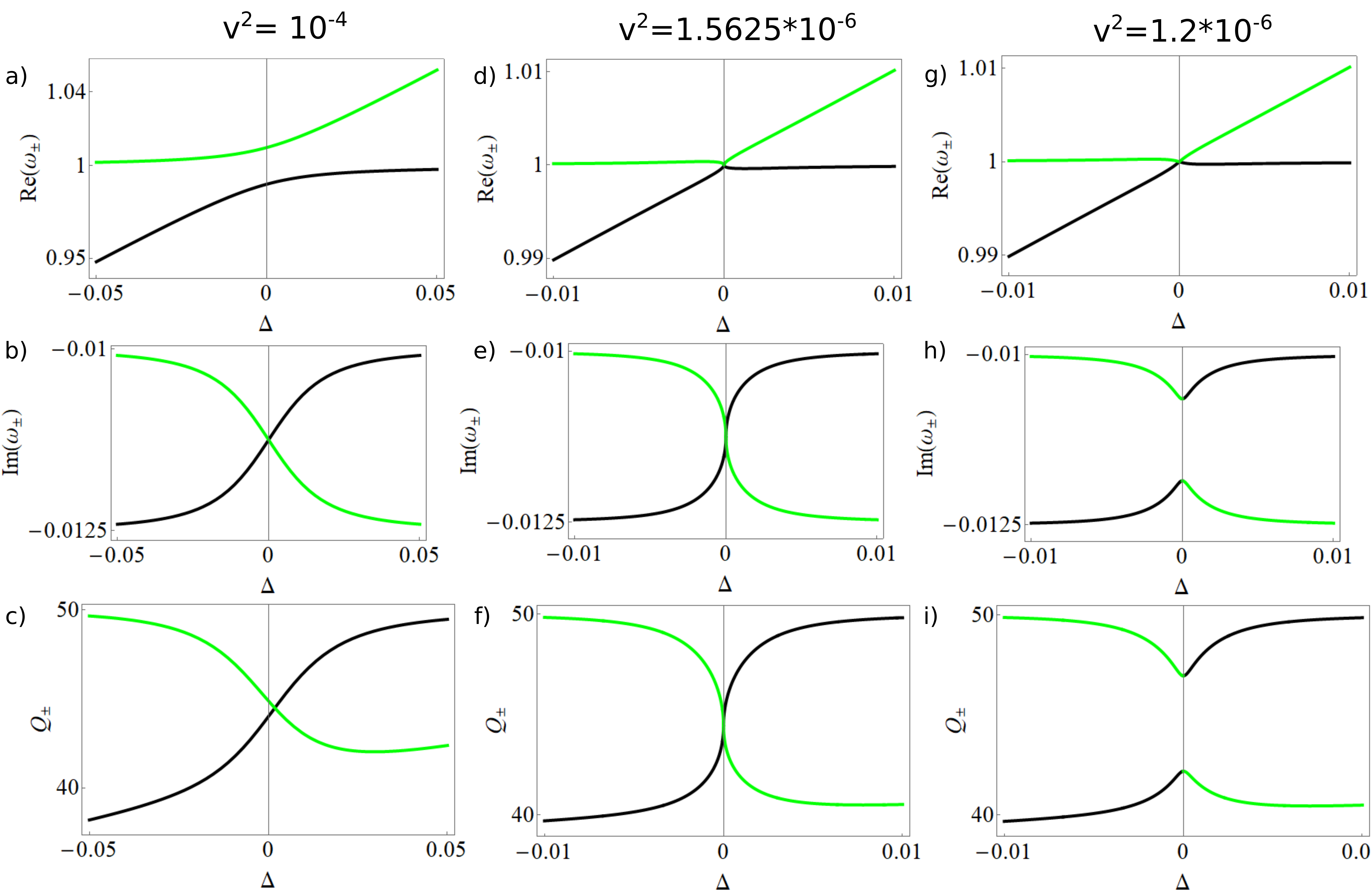} 
\caption{Real parts $\mathrm{Re}(\omega_\pm)$, imaginary parts $\mathrm{Im}(\omega_\pm)$ and Q-factors $Q_\pm$ of the two coupled frequencies $\omega_\pm$ as a function of the difference of the real parts of the uncoupled eigenfrequencies $\Delta = \mathrm{Re}(\omega_{un,2}-\omega_{un,1})$. The trajectories are given for three different couplings $v^2 = iu$, with $u$ being real and positive. For better readability a vertical line is shown at $\Delta = 0$.}
\label{FigSI_2} 
\end{figure*}

Let us first study the coupling of resonances for a coupling coefficient $v$ which is real and positive. A careful study of the behavior of the function $\gamma$ in Eq.  \eqref{def_gamma} reveals 3 different behaviors depending on the relative values of $\gamma$ and $\Delta\omega_i$. We can start by reexpressing $\gamma$ for $\omega_{\textrm{un},1}$ and $\omega_{\textrm{un},2}$ defined in Eq.~\eqref{Def_omega_un}: 
\begin{equation}
\begin{aligned}
\gamma &= \left(\frac{\Delta-i\Delta\omega_i}{2}\right)^2 + v^2\\
&= \frac{\Delta^2+4 v^2 -\Delta\omega_i^2}{4} -i\frac{\Delta*\Delta\omega_i}{2}
\end{aligned}
\end{equation}
One notices that the real part of $\gamma$ cancels out for $\Delta = \pm\sqrt{\Delta\omega_i^2- 4v^2}$ while its imaginary part cancels out for $\Delta = 0$. Studying the roots of the real part of $\gamma$, three regimes of coupling can then be distinguished depending on the relative values of $v^2$ and $\Delta\omega_i^2$: $v^2>\frac{\Delta\omega_i^2}{4}$, $v^2 = \frac{\Delta\omega_i^2}{4}$ and $v^2<\frac{\Delta\omega_i^2}{4}$. In the example we study, $\frac{\Delta\omega_i^2}{4} = 1.5625*10^{-6}$. We then plot the trajectories of the real and imaginary parts of the coupled eigenvalues for $v^2$ larger, equal and smaller than $ 1.5625*10^{-6}$. The results are plotted in the following figures along with the variation of the Q-factor as a function of $\Delta$.\\
In Fig. \ref{Fig:SI_1} a)-c), we plot these trajectories for $v^2=10^{-4}$ and thus larger than $ 1.5625*10^{-6}$. The trajectories of the eigenvalues display a behavior typical for a strong coupling as discussed in the main text with an avoided crossing of the real parts and a crossing of the imaginary parts. On both sides of this crossing, the Q-factor of one mode increases while the Q-factor of the other one decreases.\\

The trajectories of the eigenvalues when $ v^2 = 1.5625*10^{-6}$ are shown in Fig. \ref{Fig:SI_1} d)-f). It is clearly seen that the real part and the imaginary parts cross at $\Delta=0$. The two eigenfrequencies are thus completely degenerated at $\Delta =0$ which is linked to the existence of an exceptional point, a degeneracy existing in non-Hermitian systems. The behavior of the Q-factor is similar to the one observed in Fig. \ref{Fig:SI_1} c).\\
Finally, the trajectories for $v^2 < 1.5625*10^{-6}$ are shown in Fig. \ref{Fig:SI_1} g)-i). In this case, there is a crossing of the real part and an avoided crossing of the imaginary part. $\omega_{+}$ and $\omega_{-}$ are actually swapped on one side and the other of $\Delta = 0$. This behavior might certainly be seen as a jump from one Riemann sheet to the other.

\begin{figure*}
\includegraphics{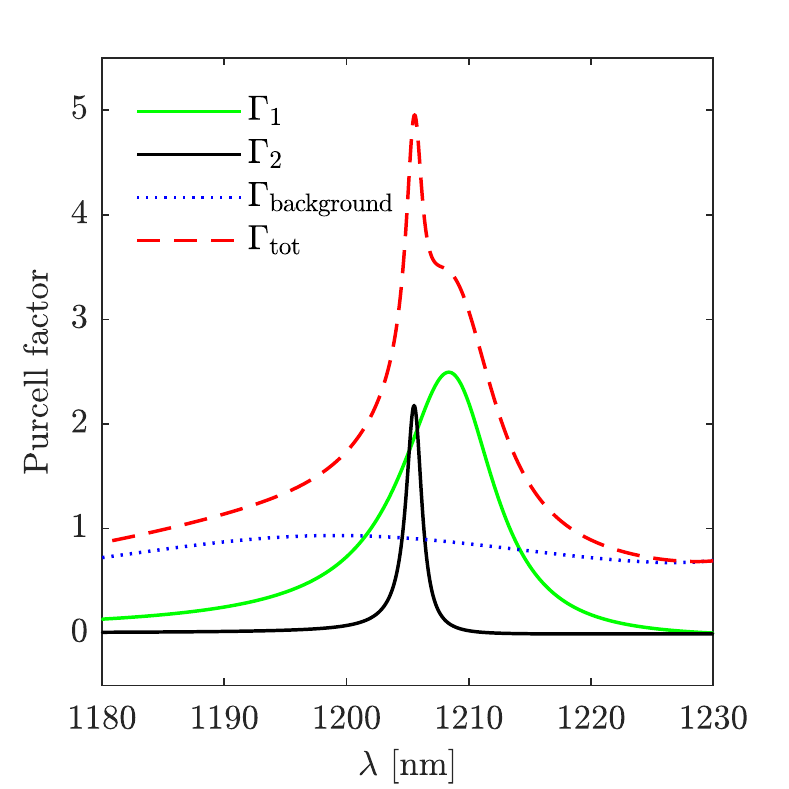} 
\caption{Purcell factor for a dipole located close to the hot spot of the low-Q mode, on the symmetry axis 700 nm above the bottom of the nanodisk. The contributions of both modes are similar and positive at this position leading to a total Purcell factor which is the superposition of the two main modes.}
\label{Fig:SI_3} 
\end{figure*}

\subsection{Coupling of resonance for $v^2$ purely imaginary} 
A  similar analysis to the one done in the previous section can be performed for $v^2$ which is purely imaginary $v^2 = i u$ with $u$ being real-valued. $\gamma$ can then be rewritten in the following way:
\begin{equation}
\begin{aligned}
\gamma &= \left(\frac{\Delta-i\Delta\omega_i}{2}\right)^2 + v^2\\
&= \frac{\Delta^2-\Delta\omega_i^2}{4} -i\frac{\Delta*\Delta\omega_i-2u}{2}
\end{aligned}
\end{equation}
This time, the roots of the real part of $\gamma$ occur for $\Delta = \pm \Delta\omega_i$ while the imaginary part of $\gamma$ vanishes for $\Delta = \frac{2u}{\Delta\omega_i}$. The different regimes of coupling now depend on the relative values of $\Delta\omega_i$ and $\frac{2u}{\Delta\omega_i}$ or equivalently the relative values of $u$ and $\frac{\Delta\omega_i^2}{2}$. We will then study the coupling of resonances when $u>\frac{\Delta\omega_i^2}{2}$, $u=\frac{\Delta\omega_i^2}{2}$ and finally $u<\frac{\Delta\omega_i^2}{2}$. In the case we study, $\frac{\Delta\omega_i^2}{2} = 3.125*10^{-6}$.\\
The trajectories of the coupled eigenvalues for $u=10^{-4}$ so larger than $\frac{\Delta\omega_i^2}{2}$ are shown in Fig.~\ref{FigSI_2} a)-c). These trajectories display a strong coupling behavior with an avoided crossing of the real parts of the coupled eigenvalues. This avoided crossing coincide this time with a peal of the imaginary part of one eigenvalue and a dip of the other one. As a consequence there is a peak of the Q-factor associated with one eigenvalue and a dip of the Q-factor of the other eigenvalue.

The trajectories of the eigenvalues for $u=3.125 10^{-6}$ which is equal to $\frac{\Delta\omega_i^2}{2}$ are shown in Fig.~\ref{FigSI_2} d)-f). This reveals the existence of an degeneracy of the eigenvalue, $i.e.$ an exceptional point, where both the real and imaginary parts of the two eigenvalues are identical.\\
Finally, when $u=2.5 10^{-6}$ which is smaller than $\frac{\Delta\omega_i^2}{2}$ the trajectories of the eigenvalues are displayed the behavior observed in Fig.~\ref{FigSI_2} g)-i). These trajectories reveal a weak coupling behavior with a crossing of the real part of the coupled eigenvalues and an anti-crossing of the imaginary part.


\section{Additional calculations for the emission of a dipole emitter}\label{appendix_sec_dipole}
In the main text we keep the position of the dipole emitter fixed and show how the coupling effects the modal contributions $\Gamma_n(\omega)$ of the Purcell enhancement at the hot spot of the high-Q mode. $\Gamma_n(\omega)$ depends on the electric field strength at the dipole position and if the quasi BIC condition is met, the contribution of the low-Q mode at this point is much smaller. For a slightly altered aspect ratio the field values become comparable and interference can be observed. Instead of varying the aspect ratio, one can also consider the enhancement at different dipole positions. For the sake of completeness in Fig.~\ref{Fig:SI_3} we provide a corresponding example where the contributions of both modes are similar. For this example the dipole is positioned at the symmetry axis close to the center of the nanodisk 700~nm above the bottom face.




\begin{thebibliography}{64}%
\makeatletter
\providecommand \@ifxundefined [1]{%
 \@ifx{#1\undefined}
}%
\providecommand \@ifnum [1]{%
 \ifnum #1\expandafter \@firstoftwo
 \else \expandafter \@secondoftwo
 \fi
}%
\providecommand \@ifx [1]{%
 \ifx #1\expandafter \@firstoftwo
 \else \expandafter \@secondoftwo
 \fi
}%
\providecommand \natexlab [1]{#1}%
\providecommand \enquote  [1]{``#1''}%
\providecommand \bibnamefont  [1]{#1}%
\providecommand \bibfnamefont [1]{#1}%
\providecommand \citenamefont [1]{#1}%
\providecommand \href@noop [0]{\@secondoftwo}%
\providecommand \href [0]{\begingroup \@sanitize@url \@href}%
\providecommand \@href[1]{\@@startlink{#1}\@@href}%
\providecommand \@@href[1]{\endgroup#1\@@endlink}%
\providecommand \@sanitize@url [0]{\catcode `\\12\catcode `\$12\catcode
  `\&12\catcode `\#12\catcode `\^12\catcode `\_12\catcode `\%12\relax}%
\providecommand \@@startlink[1]{}%
\providecommand \@@endlink[0]{}%
\providecommand \url  [0]{\begingroup\@sanitize@url \@url }%
\providecommand \@url [1]{\endgroup\@href {#1}{\urlprefix }}%
\providecommand \urlprefix  [0]{URL }%
\providecommand \Eprint [0]{\href }%
\providecommand \doibase [0]{https://doi.org/}%
\providecommand \selectlanguage [0]{\@gobble}%
\providecommand \bibinfo  [0]{\@secondoftwo}%
\providecommand \bibfield  [0]{\@secondoftwo}%
\providecommand \translation [1]{[#1]}%
\providecommand \BibitemOpen [0]{}%
\providecommand \bibitemStop [0]{}%
\providecommand \bibitemNoStop [0]{.\EOS\space}%
\providecommand \EOS [0]{\spacefactor3000\relax}%
\providecommand \BibitemShut  [1]{\csname bibitem#1\endcsname}%
\let\auto@bib@innerbib\@empty
\bibitem [{\citenamefont {Maier}\ and\ \citenamefont
  {Atwater}(2005)}]{maier2005}%
  \BibitemOpen
  \bibfield  {author} {\bibinfo {author} {\bibfnamefont {S.~A.}\ \bibnamefont
  {Maier}}\ and\ \bibinfo {author} {\bibfnamefont {H.~A.}\ \bibnamefont
  {Atwater}},\ }\bibfield  {title} {\bibinfo {title} {Plasmonics:
  {L}ocalization and guiding of electromagnetic energy in metal/dielectric
  structures},\ }\href {https://doi.org/10.1063/1.1951057} {\bibfield
  {journal} {\bibinfo  {journal} {J. Appl. Phys.}\ }\textbf {\bibinfo {volume}
  {98}},\ \bibinfo {pages} {011101} (\bibinfo {year} {2005})}\BibitemShut
  {NoStop}%
\bibitem [{\citenamefont {Maier}(2007)}]{maier2007}%
  \BibitemOpen
  \bibfield  {author} {\bibinfo {author} {\bibfnamefont {S.~A.}\ \bibnamefont
  {Maier}},\ }\href {https://doi.org/10.1007/0-387-37825-1} {\emph {\bibinfo
  {title} {Plasmonics: {F}undamentals and Applications}}}\ (\bibinfo
  {publisher} {Springer Science \& Business Media},\ \bibinfo {year}
  {2007})\BibitemShut {NoStop}%
\bibitem [{\citenamefont {Novotny}\ and\ \citenamefont
  {Van~Hulst}(2011)}]{novotny2011}%
  \BibitemOpen
  \bibfield  {author} {\bibinfo {author} {\bibfnamefont {L.}~\bibnamefont
  {Novotny}}\ and\ \bibinfo {author} {\bibfnamefont {N.}~\bibnamefont
  {Van~Hulst}},\ }\bibfield  {title} {\bibinfo {title} {Antennas for light},\
  }\href {https://doi.org/10.1038/nphoton.2010.237} {\bibfield  {journal}
  {\bibinfo  {journal} {Nat. Photonics}\ }\textbf {\bibinfo {volume} {5}},\
  \bibinfo {pages} {83} (\bibinfo {year} {2011})}\BibitemShut {NoStop}%
\bibitem [{\citenamefont {Novotny}\ and\ \citenamefont
  {Hecht}(2012)}]{novotny2012}%
  \BibitemOpen
  \bibfield  {author} {\bibinfo {author} {\bibfnamefont {L.}~\bibnamefont
  {Novotny}}\ and\ \bibinfo {author} {\bibfnamefont {B.}~\bibnamefont
  {Hecht}},\ }\href {https://doi.org/10.1017/CBO9780511794193} {\emph {\bibinfo
  {title} {Principles of nano-optics}}}\ (\bibinfo  {publisher} {Cambridge
  University Press},\ \bibinfo {year} {2012})\BibitemShut {NoStop}%
\bibitem [{\citenamefont {Evlyukhin}\ \emph {et~al.}(2010)\citenamefont
  {Evlyukhin}, \citenamefont {Reinhardt}, \citenamefont {Seidel}, \citenamefont
  {Luk'yanchuk},\ and\ \citenamefont {Chichkov}}]{Evlyukhin_2010}%
  \BibitemOpen
  \bibfield  {author} {\bibinfo {author} {\bibfnamefont {A.~B.}\ \bibnamefont
  {Evlyukhin}}, \bibinfo {author} {\bibfnamefont {C.}~\bibnamefont
  {Reinhardt}}, \bibinfo {author} {\bibfnamefont {A.}~\bibnamefont {Seidel}},
  \bibinfo {author} {\bibfnamefont {B.~S.}\ \bibnamefont {Luk'yanchuk}},\ and\
  \bibinfo {author} {\bibfnamefont {B.~N.}\ \bibnamefont {Chichkov}},\
  }\bibfield  {title} {\bibinfo {title} {Optical response features of
  {Si}-nanoparticle arrays},\ }\href
  {https://doi.org/10.1103/PhysRevB.82.045404} {\bibfield  {journal} {\bibinfo
  {journal} {Phys. Rev. B}\ }\textbf {\bibinfo {volume} {82}},\ \bibinfo
  {pages} {045404} (\bibinfo {year} {2010})}\BibitemShut {NoStop}%
\bibitem [{\citenamefont {Garc\'{i}a-Etxarri}\ \emph
  {et~al.}(2011)\citenamefont {Garc\'{i}a-Etxarri}, \citenamefont
  {G\'{o}mez-Medina}, \citenamefont {Froufe-P\'{e}rez}, \citenamefont
  {L\'{o}pez}, \citenamefont {Chantada}, \citenamefont {Scheffold},
  \citenamefont {Aizpurua}, \citenamefont {Nieto-Vesperinas},\ and\
  \citenamefont {S\'{a}enz}}]{Garcia-Etxarri_2011}%
  \BibitemOpen
  \bibfield  {author} {\bibinfo {author} {\bibfnamefont {A.}~\bibnamefont
  {Garc\'{i}a-Etxarri}}, \bibinfo {author} {\bibfnamefont {R.}~\bibnamefont
  {G\'{o}mez-Medina}}, \bibinfo {author} {\bibfnamefont {L.~S.}\ \bibnamefont
  {Froufe-P\'{e}rez}}, \bibinfo {author} {\bibfnamefont {C.}~\bibnamefont
  {L\'{o}pez}}, \bibinfo {author} {\bibfnamefont {L.}~\bibnamefont {Chantada}},
  \bibinfo {author} {\bibfnamefont {F.}~\bibnamefont {Scheffold}}, \bibinfo
  {author} {\bibfnamefont {J.}~\bibnamefont {Aizpurua}}, \bibinfo {author}
  {\bibfnamefont {M.}~\bibnamefont {Nieto-Vesperinas}},\ and\ \bibinfo {author}
  {\bibfnamefont {J.~J.}\ \bibnamefont {S\'{a}enz}},\ }\bibfield  {title}
  {\bibinfo {title} {Strong magnetic response of submicron silicon particles in
  the infrared},\ }\href {https://doi.org/10.1364/OE.19.004815} {\bibfield
  {journal} {\bibinfo  {journal} {Opt. Express}\ }\textbf {\bibinfo {volume}
  {19}},\ \bibinfo {pages} {4815} (\bibinfo {year} {2011})}\BibitemShut
  {NoStop}%
\bibitem [{\citenamefont {Kuznetsov}\ \emph {et~al.}(2016)\citenamefont
  {Kuznetsov}, \citenamefont {Miroshnichenko}, \citenamefont {Brongersma},
  \citenamefont {Kivshar},\ and\ \citenamefont
  {Luk’yanchuk}}]{kuznetsov2016}%
  \BibitemOpen
  \bibfield  {author} {\bibinfo {author} {\bibfnamefont {A.~I.}\ \bibnamefont
  {Kuznetsov}}, \bibinfo {author} {\bibfnamefont {A.~E.}\ \bibnamefont
  {Miroshnichenko}}, \bibinfo {author} {\bibfnamefont {M.~L.}\ \bibnamefont
  {Brongersma}}, \bibinfo {author} {\bibfnamefont {Y.~S.}\ \bibnamefont
  {Kivshar}},\ and\ \bibinfo {author} {\bibfnamefont {B.}~\bibnamefont
  {Luk’yanchuk}},\ }\bibfield  {title} {\bibinfo {title} {Optically resonant
  dielectric nanostructures},\ }\href {https://doi.org/10.1126/science.aag2472}
  {\bibfield  {journal} {\bibinfo  {journal} {Science}\ }\textbf {\bibinfo
  {volume} {354}},\ \bibinfo {pages} {aag2472} (\bibinfo {year}
  {2016})}\BibitemShut {NoStop}%
\bibitem [{\citenamefont {Barreda}\ \emph {et~al.}(2019)\citenamefont
  {Barreda}, \citenamefont {Saiz}, \citenamefont {González}, \citenamefont
  {Moreno},\ and\ \citenamefont {Albella}}]{Barreda_AIP_2019}%
  \BibitemOpen
  \bibfield  {author} {\bibinfo {author} {\bibfnamefont {A.~I.}\ \bibnamefont
  {Barreda}}, \bibinfo {author} {\bibfnamefont {J.~M.}\ \bibnamefont {Saiz}},
  \bibinfo {author} {\bibfnamefont {F.}~\bibnamefont {González}}, \bibinfo
  {author} {\bibfnamefont {F.}~\bibnamefont {Moreno}},\ and\ \bibinfo {author}
  {\bibfnamefont {P.}~\bibnamefont {Albella}},\ }\bibfield  {title} {\bibinfo
  {title} {Recent advances in high refractive index dielectric nanoantennas:
  {B}asics and applications},\ }\href {https://doi.org/10.1063/1.5087402}
  {\bibfield  {journal} {\bibinfo  {journal} {AIP Adv.}\ }\textbf {\bibinfo
  {volume} {9}},\ \bibinfo {pages} {040701} (\bibinfo {year}
  {2019})}\BibitemShut {NoStop}%
\bibitem [{\citenamefont {Aharonovich}\ \emph {et~al.}(2011)\citenamefont
  {Aharonovich}, \citenamefont {Castelletto}, \citenamefont {Simpson},
  \citenamefont {Su}, \citenamefont {Greentree},\ and\ \citenamefont
  {Prawer}}]{Aharonovich2011}%
  \BibitemOpen
  \bibfield  {author} {\bibinfo {author} {\bibfnamefont {I.}~\bibnamefont
  {Aharonovich}}, \bibinfo {author} {\bibfnamefont {S.}~\bibnamefont
  {Castelletto}}, \bibinfo {author} {\bibfnamefont {D.~A.}\ \bibnamefont
  {Simpson}}, \bibinfo {author} {\bibfnamefont {C.-H.}\ \bibnamefont {Su}},
  \bibinfo {author} {\bibfnamefont {A.~D.}\ \bibnamefont {Greentree}},\ and\
  \bibinfo {author} {\bibfnamefont {S.}~\bibnamefont {Prawer}},\ }\bibfield
  {title} {\bibinfo {title} {Diamond-based single-photon emitters},\ }\href
  {https://doi.org/10.1088/0034-4885/74/7/076501} {\bibfield  {journal}
  {\bibinfo  {journal} {Rep. Prog. Phys.}\ }\textbf {\bibinfo {volume} {74}},\
  \bibinfo {pages} {076501} (\bibinfo {year} {2011})}\BibitemShut {NoStop}%
\bibitem [{\citenamefont {Lodahl}\ \emph {et~al.}(2015)\citenamefont {Lodahl},
  \citenamefont {Mahmoodian},\ and\ \citenamefont {Stobbe}}]{Lodahl2015}%
  \BibitemOpen
  \bibfield  {author} {\bibinfo {author} {\bibfnamefont {P.}~\bibnamefont
  {Lodahl}}, \bibinfo {author} {\bibfnamefont {S.}~\bibnamefont {Mahmoodian}},\
  and\ \bibinfo {author} {\bibfnamefont {S.}~\bibnamefont {Stobbe}},\
  }\bibfield  {title} {\bibinfo {title} {Interfacing single photons and single
  quantum dots with photonic nanostructures},\ }\href
  {https://doi.org/10.1103/RevModPhys.87.347} {\bibfield  {journal} {\bibinfo
  {journal} {Rev. Mod. Phys.}\ }\textbf {\bibinfo {volume} {87}},\ \bibinfo
  {pages} {347} (\bibinfo {year} {2015})}\BibitemShut {NoStop}%
\bibitem [{\citenamefont {Purcell}(1946)}]{Purcell_1946}%
  \BibitemOpen
  \bibfield  {author} {\bibinfo {author} {\bibfnamefont {E.~M.}\ \bibnamefont
  {Purcell}},\ }\bibfield  {title} {\bibinfo {title} {{Spontaneous emission
  probabilities at radio frequencies}},\ }\href
  {https://doi.org/10.1103/PhysRev.69.674} {\bibfield  {journal} {\bibinfo
  {journal} {Phys. Rev.}\ }\textbf {\bibinfo {volume} {69}},\ \bibinfo {pages}
  {681} (\bibinfo {year} {1946})}\BibitemShut {NoStop}%
\bibitem [{\citenamefont {Drexhage}(1970)}]{drexhage1970}%
  \BibitemOpen
  \bibfield  {author} {\bibinfo {author} {\bibfnamefont {K.}~\bibnamefont
  {Drexhage}},\ }\bibfield  {title} {\bibinfo {title} {Influence of a
  dielectric interface on fluorescence decay time},\ }\href
  {https://doi.org/10.1016/0022-2313(70)90082-7} {\bibfield  {journal}
  {\bibinfo  {journal} {J. Lumin.}\ }\textbf {\bibinfo {volume} {1}},\ \bibinfo
  {pages} {693} (\bibinfo {year} {1970})}\BibitemShut {NoStop}%
\bibitem [{\citenamefont {Holsteen}\ \emph {et~al.}(2017)\citenamefont
  {Holsteen}, \citenamefont {Raza}, \citenamefont {Fan}, \citenamefont {Kik},\
  and\ \citenamefont {Brongersma}}]{holsteen2017}%
  \BibitemOpen
  \bibfield  {author} {\bibinfo {author} {\bibfnamefont {A.~L.}\ \bibnamefont
  {Holsteen}}, \bibinfo {author} {\bibfnamefont {S.}~\bibnamefont {Raza}},
  \bibinfo {author} {\bibfnamefont {P.}~\bibnamefont {Fan}}, \bibinfo {author}
  {\bibfnamefont {P.~G.}\ \bibnamefont {Kik}},\ and\ \bibinfo {author}
  {\bibfnamefont {M.~L.}\ \bibnamefont {Brongersma}},\ }\bibfield  {title}
  {\bibinfo {title} {Purcell effect for active tuning of light scattering from
  semiconductor optical antennas},\ }\href
  {https://doi.org/10.1126/science.aao5371} {\bibfield  {journal} {\bibinfo
  {journal} {Science}\ }\textbf {\bibinfo {volume} {358}},\ \bibinfo {pages}
  {1407} (\bibinfo {year} {2017})}\BibitemShut {NoStop}%
\bibitem [{\citenamefont {Albella}\ \emph {et~al.}(2013)\citenamefont
  {Albella}, \citenamefont {Poyli}, \citenamefont {Schmidt}, \citenamefont
  {Maier}, \citenamefont {Moreno}, \citenamefont {Sáenz},\ and\ \citenamefont
  {Aizpurua}}]{Albella_2013}%
  \BibitemOpen
  \bibfield  {author} {\bibinfo {author} {\bibfnamefont {P.}~\bibnamefont
  {Albella}}, \bibinfo {author} {\bibfnamefont {M.~A.}\ \bibnamefont {Poyli}},
  \bibinfo {author} {\bibfnamefont {M.~K.}\ \bibnamefont {Schmidt}}, \bibinfo
  {author} {\bibfnamefont {S.~A.}\ \bibnamefont {Maier}}, \bibinfo {author}
  {\bibfnamefont {F.}~\bibnamefont {Moreno}}, \bibinfo {author} {\bibfnamefont
  {J.~J.}\ \bibnamefont {Sáenz}},\ and\ \bibinfo {author} {\bibfnamefont
  {J.}~\bibnamefont {Aizpurua}},\ }\bibfield  {title} {\bibinfo {title}
  {Low-loss electric and magnetic field-enhanced spectroscopy with
  subwavelength silicon dimers},\ }\href {https://doi.org/10.1021/jp4027018}
  {\bibfield  {journal} {\bibinfo  {journal} {J. Phys. Chem. C}\ }\textbf
  {\bibinfo {volume} {117}},\ \bibinfo {pages} {13573} (\bibinfo {year}
  {2013})}\BibitemShut {NoStop}%
\bibitem [{\citenamefont {Zambrana-Puyalto}\ and\ \citenamefont
  {Bonod}(2015)}]{zambrana2015}%
  \BibitemOpen
  \bibfield  {author} {\bibinfo {author} {\bibfnamefont {X.}~\bibnamefont
  {Zambrana-Puyalto}}\ and\ \bibinfo {author} {\bibfnamefont {N.}~\bibnamefont
  {Bonod}},\ }\bibfield  {title} {\bibinfo {title} {Purcell factor of spherical
  {M}ie resonators},\ }\href {https://doi.org/10.1103/PhysRevB.91.195422}
  {\bibfield  {journal} {\bibinfo  {journal} {Phys. Rev. B}\ }\textbf {\bibinfo
  {volume} {91}},\ \bibinfo {pages} {195422} (\bibinfo {year}
  {2015})}\BibitemShut {NoStop}%
\bibitem [{\citenamefont {Rolly}\ \emph {et~al.}(2012)\citenamefont {Rolly},
  \citenamefont {Bebey}, \citenamefont {Bidault}, \citenamefont {Stout},\ and\
  \citenamefont {Bonod}}]{rolly2012}%
  \BibitemOpen
  \bibfield  {author} {\bibinfo {author} {\bibfnamefont {B.}~\bibnamefont
  {Rolly}}, \bibinfo {author} {\bibfnamefont {B.}~\bibnamefont {Bebey}},
  \bibinfo {author} {\bibfnamefont {S.}~\bibnamefont {Bidault}}, \bibinfo
  {author} {\bibfnamefont {B.}~\bibnamefont {Stout}},\ and\ \bibinfo {author}
  {\bibfnamefont {N.}~\bibnamefont {Bonod}},\ }\bibfield  {title} {\bibinfo
  {title} {Promoting magnetic dipolar transition in trivalent lanthanide ions
  with lossless {M}ie resonances},\ }\href
  {https://doi.org/10.1103/PhysRevB.85.245432} {\bibfield  {journal} {\bibinfo
  {journal} {Phy. Rev. B}\ }\textbf {\bibinfo {volume} {85}},\ \bibinfo {pages}
  {245432} (\bibinfo {year} {2012})}\BibitemShut {NoStop}%
\bibitem [{\citenamefont {Schmidt}\ \emph {et~al.}(2012)\citenamefont
  {Schmidt}, \citenamefont {Esteban}, \citenamefont {S{\'a}enz}, \citenamefont
  {Su{\'a}rez-Lacalle}, \citenamefont {Mackowski},\ and\ \citenamefont
  {Aizpurua}}]{schmidt2012}%
  \BibitemOpen
  \bibfield  {author} {\bibinfo {author} {\bibfnamefont {M.~K.}\ \bibnamefont
  {Schmidt}}, \bibinfo {author} {\bibfnamefont {R.}~\bibnamefont {Esteban}},
  \bibinfo {author} {\bibfnamefont {J.}~\bibnamefont {S{\'a}enz}}, \bibinfo
  {author} {\bibfnamefont {I.}~\bibnamefont {Su{\'a}rez-Lacalle}}, \bibinfo
  {author} {\bibfnamefont {S.}~\bibnamefont {Mackowski}},\ and\ \bibinfo
  {author} {\bibfnamefont {J.}~\bibnamefont {Aizpurua}},\ }\bibfield  {title}
  {\bibinfo {title} {Dielectric antennas -- a suitable platform for controlling
  magnetic dipolar emission},\ }\href {https://doi.org/10.1364/OE.20.013636}
  {\bibfield  {journal} {\bibinfo  {journal} {Opt. Express}\ }\textbf {\bibinfo
  {volume} {20}},\ \bibinfo {pages} {13636} (\bibinfo {year}
  {2012})}\BibitemShut {NoStop}%
\bibitem [{\citenamefont {Sanz-Paz}\ \emph {et~al.}(2018)\citenamefont
  {Sanz-Paz}, \citenamefont {Ernandes}, \citenamefont {Esparza}, \citenamefont
  {Burr}, \citenamefont {van Hulst}, \citenamefont {Maitre}, \citenamefont
  {Aigouy}, \citenamefont {Gacoin}, \citenamefont {Bonod}, \citenamefont
  {Garcia-Parajo} \emph {et~al.}}]{sanz2018}%
  \BibitemOpen
  \bibfield  {author} {\bibinfo {author} {\bibfnamefont {M.}~\bibnamefont
  {Sanz-Paz}}, \bibinfo {author} {\bibfnamefont {C.}~\bibnamefont {Ernandes}},
  \bibinfo {author} {\bibfnamefont {J.~U.}\ \bibnamefont {Esparza}}, \bibinfo
  {author} {\bibfnamefont {G.~W.}\ \bibnamefont {Burr}}, \bibinfo {author}
  {\bibfnamefont {N.~F.}\ \bibnamefont {van Hulst}}, \bibinfo {author}
  {\bibfnamefont {A.}~\bibnamefont {Maitre}}, \bibinfo {author} {\bibfnamefont
  {L.}~\bibnamefont {Aigouy}}, \bibinfo {author} {\bibfnamefont
  {T.}~\bibnamefont {Gacoin}}, \bibinfo {author} {\bibfnamefont
  {N.}~\bibnamefont {Bonod}}, \bibinfo {author} {\bibfnamefont {M.~F.}\
  \bibnamefont {Garcia-Parajo}}, \emph {et~al.},\ }\bibfield  {title} {\bibinfo
  {title} {Enhancing magnetic light emission with all-dielectric optical
  nanoantennas},\ }\href {https://doi.org/10.1021/acs.nanolett.8b00548}
  {\bibfield  {journal} {\bibinfo  {journal} {Nano Lett.}\ }\textbf {\bibinfo
  {volume} {18}},\ \bibinfo {pages} {3481} (\bibinfo {year}
  {2018})}\BibitemShut {NoStop}%
\bibitem [{\citenamefont {Vaskin}\ \emph {et~al.}(2019)\citenamefont {Vaskin},
  \citenamefont {Mashhadi}, \citenamefont {Steinert}, \citenamefont {Chong},
  \citenamefont {Keene}, \citenamefont {Nanz}, \citenamefont {Abass},
  \citenamefont {Rusak}, \citenamefont {Choi}, \citenamefont
  {Fernandez-Corbaton} \emph {et~al.}}]{vaskin2019}%
  \BibitemOpen
  \bibfield  {author} {\bibinfo {author} {\bibfnamefont {A.}~\bibnamefont
  {Vaskin}}, \bibinfo {author} {\bibfnamefont {S.}~\bibnamefont {Mashhadi}},
  \bibinfo {author} {\bibfnamefont {M.}~\bibnamefont {Steinert}}, \bibinfo
  {author} {\bibfnamefont {K.~E.}\ \bibnamefont {Chong}}, \bibinfo {author}
  {\bibfnamefont {D.}~\bibnamefont {Keene}}, \bibinfo {author} {\bibfnamefont
  {S.}~\bibnamefont {Nanz}}, \bibinfo {author} {\bibfnamefont {A.}~\bibnamefont
  {Abass}}, \bibinfo {author} {\bibfnamefont {E.}~\bibnamefont {Rusak}},
  \bibinfo {author} {\bibfnamefont {D.-Y.}\ \bibnamefont {Choi}}, \bibinfo
  {author} {\bibfnamefont {I.}~\bibnamefont {Fernandez-Corbaton}}, \emph
  {et~al.},\ }\bibfield  {title} {\bibinfo {title} {Manipulation of magnetic
  dipole emission from {Eu3+} with {M}ie-resonant dielectric metasurfaces},\
  }\href {https://doi.org/10.1021/acs.nanolett.8b04268} {\bibfield  {journal}
  {\bibinfo  {journal} {Nano Lett.}\ }\textbf {\bibinfo {volume} {19}},\
  \bibinfo {pages} {1015} (\bibinfo {year} {2019})}\BibitemShut {NoStop}%
\bibitem [{\citenamefont {Sugimoto}\ and\ \citenamefont
  {Fujii}(2021)}]{sugimoto2021}%
  \BibitemOpen
  \bibfield  {author} {\bibinfo {author} {\bibfnamefont {H.}~\bibnamefont
  {Sugimoto}}\ and\ \bibinfo {author} {\bibfnamefont {M.}~\bibnamefont
  {Fujii}},\ }\bibfield  {title} {\bibinfo {title} {Magnetic {P}urcell
  enhancement by magnetic quadrupole resonance of dielectric nanosphere
  antenna},\ }\href {https://doi.org/10.1021/acsphotonics.1c00375} {\bibfield
  {journal} {\bibinfo  {journal} {ACS Photonics}\ }\textbf {\bibinfo {volume}
  {8}},\ \bibinfo {pages} {1794–1800} (\bibinfo {year} {2021})}\BibitemShut
  {NoStop}%
\bibitem [{\citenamefont {Karaveli}\ and\ \citenamefont
  {Zia}(2011)}]{karaveli2011}%
  \BibitemOpen
  \bibfield  {author} {\bibinfo {author} {\bibfnamefont {S.}~\bibnamefont
  {Karaveli}}\ and\ \bibinfo {author} {\bibfnamefont {R.}~\bibnamefont {Zia}},\
  }\bibfield  {title} {\bibinfo {title} {Spectral tuning by selective
  enhancement of electric and magnetic dipole emission},\ }\href
  {https://doi.org/10.1103/PhysRevLett.106.193004} {\bibfield  {journal}
  {\bibinfo  {journal} {Phys. Rev. Lett.}\ }\textbf {\bibinfo {volume} {106}},\
  \bibinfo {pages} {193004} (\bibinfo {year} {2011})}\BibitemShut {NoStop}%
\bibitem [{\citenamefont {Baranov}\ \emph {et~al.}(2017)\citenamefont
  {Baranov}, \citenamefont {Savelev}, \citenamefont {Li}, \citenamefont
  {Krasnok},\ and\ \citenamefont {Al{\`u}}}]{baranov2017}%
  \BibitemOpen
  \bibfield  {author} {\bibinfo {author} {\bibfnamefont {D.~G.}\ \bibnamefont
  {Baranov}}, \bibinfo {author} {\bibfnamefont {R.~S.}\ \bibnamefont
  {Savelev}}, \bibinfo {author} {\bibfnamefont {S.~V.}\ \bibnamefont {Li}},
  \bibinfo {author} {\bibfnamefont {A.~E.}\ \bibnamefont {Krasnok}},\ and\
  \bibinfo {author} {\bibfnamefont {A.}~\bibnamefont {Al{\`u}}},\ }\bibfield
  {title} {\bibinfo {title} {Modifying magnetic dipole spontaneous emission
  with nanophotonic structures},\ }\href
  {https://doi.org/10.1002/lpor.201600268} {\bibfield  {journal} {\bibinfo
  {journal} {Laser Photonics Rev.}\ }\textbf {\bibinfo {volume} {11}},\
  \bibinfo {pages} {1600268} (\bibinfo {year} {2017})}\BibitemShut {NoStop}%
\bibitem [{\citenamefont {Rutckaia}\ \emph {et~al.}(2017)\citenamefont
  {Rutckaia}, \citenamefont {Heyroth}, \citenamefont {Novikov}, \citenamefont
  {Shaleev}, \citenamefont {Petrov},\ and\ \citenamefont
  {Schilling}}]{rutckaia2017}%
  \BibitemOpen
  \bibfield  {author} {\bibinfo {author} {\bibfnamefont {V.}~\bibnamefont
  {Rutckaia}}, \bibinfo {author} {\bibfnamefont {F.}~\bibnamefont {Heyroth}},
  \bibinfo {author} {\bibfnamefont {A.}~\bibnamefont {Novikov}}, \bibinfo
  {author} {\bibfnamefont {M.}~\bibnamefont {Shaleev}}, \bibinfo {author}
  {\bibfnamefont {M.}~\bibnamefont {Petrov}},\ and\ \bibinfo {author}
  {\bibfnamefont {J.}~\bibnamefont {Schilling}},\ }\bibfield  {title} {\bibinfo
  {title} {Quantum dot emission driven by {M}ie resonances in silicon
  nanostructures},\ }\href {https://doi.org/10.1021/acs.nanolett.7b03248}
  {\bibfield  {journal} {\bibinfo  {journal} {Nano Lett.}\ }\textbf {\bibinfo
  {volume} {17}},\ \bibinfo {pages} {6886} (\bibinfo {year}
  {2017})}\BibitemShut {NoStop}%
\bibitem [{\citenamefont {Barreda}\ \emph {et~al.}(2021)\citenamefont
  {Barreda}, \citenamefont {Hell}, \citenamefont {Weissflog}, \citenamefont
  {Minovich}, \citenamefont {Pertsch},\ and\ \citenamefont
  {Staude}}]{barreda2021}%
  \BibitemOpen
  \bibfield  {author} {\bibinfo {author} {\bibfnamefont {A.}~\bibnamefont
  {Barreda}}, \bibinfo {author} {\bibfnamefont {S.}~\bibnamefont {Hell}},
  \bibinfo {author} {\bibfnamefont {M.}~\bibnamefont {Weissflog}}, \bibinfo
  {author} {\bibfnamefont {A.}~\bibnamefont {Minovich}}, \bibinfo {author}
  {\bibfnamefont {T.}~\bibnamefont {Pertsch}},\ and\ \bibinfo {author}
  {\bibfnamefont {I.}~\bibnamefont {Staude}},\ }\bibfield  {title} {\bibinfo
  {title} {Metal, dielectric and hybrid nanoantennas for enhancing the emission
  of single quantum dots: {A} comparative study},\ }\href
  {https://doi.org/10.1016/j.jqsrt.2021.107900} {\bibfield  {journal} {\bibinfo
   {journal} {J. Quant. Spectrosc. Radiat. Transf.}\ }\textbf {\bibinfo
  {volume} {276}},\ \bibinfo {pages} {107900} (\bibinfo {year}
  {2021})}\BibitemShut {NoStop}%
\bibitem [{\citenamefont {Casabone}\ \emph {et~al.}(2021)\citenamefont
  {Casabone}, \citenamefont {Deshmukh}, \citenamefont {Liu}, \citenamefont
  {Serrano}, \citenamefont {Ferrier}, \citenamefont {H{\"u}mmer}, \citenamefont
  {Goldner}, \citenamefont {Hunger},\ and\ \citenamefont
  {de~Riedmatten}}]{casabone2021}%
  \BibitemOpen
  \bibfield  {author} {\bibinfo {author} {\bibfnamefont {B.}~\bibnamefont
  {Casabone}}, \bibinfo {author} {\bibfnamefont {C.}~\bibnamefont {Deshmukh}},
  \bibinfo {author} {\bibfnamefont {S.}~\bibnamefont {Liu}}, \bibinfo {author}
  {\bibfnamefont {D.}~\bibnamefont {Serrano}}, \bibinfo {author} {\bibfnamefont
  {A.}~\bibnamefont {Ferrier}}, \bibinfo {author} {\bibfnamefont
  {T.}~\bibnamefont {H{\"u}mmer}}, \bibinfo {author} {\bibfnamefont
  {P.}~\bibnamefont {Goldner}}, \bibinfo {author} {\bibfnamefont
  {D.}~\bibnamefont {Hunger}},\ and\ \bibinfo {author} {\bibfnamefont
  {H.}~\bibnamefont {de~Riedmatten}},\ }\bibfield  {title} {\bibinfo {title}
  {Dynamic control of {P}urcell enhanced emission of erbium ions in
  nanoparticles},\ }\href {https://doi.org/10.1038/s41467-021-23632-9}
  {\bibfield  {journal} {\bibinfo  {journal} {Nat. Commun.}\ }\textbf {\bibinfo
  {volume} {12}},\ \bibinfo {pages} {3570} (\bibinfo {year}
  {2021})}\BibitemShut {NoStop}%
\bibitem [{\citenamefont {Zalogina}\ \emph {et~al.}(2018)\citenamefont
  {Zalogina}, \citenamefont {Savelev}, \citenamefont {Ushakova}, \citenamefont
  {Zograf}, \citenamefont {Komissarenko}, \citenamefont {Milichko},
  \citenamefont {Makarov}, \citenamefont {Zuev},\ and\ \citenamefont
  {Shadrivov}}]{zalogina2018}%
  \BibitemOpen
  \bibfield  {author} {\bibinfo {author} {\bibfnamefont {A.~S.}\ \bibnamefont
  {Zalogina}}, \bibinfo {author} {\bibfnamefont {R.}~\bibnamefont {Savelev}},
  \bibinfo {author} {\bibfnamefont {E.~V.}\ \bibnamefont {Ushakova}}, \bibinfo
  {author} {\bibfnamefont {G.}~\bibnamefont {Zograf}}, \bibinfo {author}
  {\bibfnamefont {F.}~\bibnamefont {Komissarenko}}, \bibinfo {author}
  {\bibfnamefont {V.}~\bibnamefont {Milichko}}, \bibinfo {author}
  {\bibfnamefont {S.}~\bibnamefont {Makarov}}, \bibinfo {author} {\bibfnamefont
  {D.}~\bibnamefont {Zuev}},\ and\ \bibinfo {author} {\bibfnamefont
  {I.}~\bibnamefont {Shadrivov}},\ }\bibfield  {title} {\bibinfo {title}
  {Purcell effect in active diamond nanoantennas},\ }\href
  {https://doi.org/10.1039/C7NR07953B} {\bibfield  {journal} {\bibinfo
  {journal} {Nanoscale}\ }\textbf {\bibinfo {volume} {10}},\ \bibinfo {pages}
  {8721} (\bibinfo {year} {2018})}\BibitemShut {NoStop}%
\bibitem [{\citenamefont {Hsu}\ \emph {et~al.}(2016)\citenamefont {Hsu},
  \citenamefont {Zhen}, \citenamefont {Stone}, \citenamefont {Joannopoulos},\
  and\ \citenamefont {Solja{\v{c}}i{\'c}}}]{hsu2016}%
  \BibitemOpen
  \bibfield  {author} {\bibinfo {author} {\bibfnamefont {C.~W.}\ \bibnamefont
  {Hsu}}, \bibinfo {author} {\bibfnamefont {B.}~\bibnamefont {Zhen}}, \bibinfo
  {author} {\bibfnamefont {A.~D.}\ \bibnamefont {Stone}}, \bibinfo {author}
  {\bibfnamefont {J.~D.}\ \bibnamefont {Joannopoulos}},\ and\ \bibinfo {author}
  {\bibfnamefont {M.}~\bibnamefont {Solja{\v{c}}i{\'c}}},\ }\bibfield  {title}
  {\bibinfo {title} {Bound states in the continuum},\ }\href
  {https://doi.org/10.1038/natrevmats.2016.48} {\bibfield  {journal} {\bibinfo
  {journal} {Nat. Rev. Mater.}\ }\textbf {\bibinfo {volume} {1}},\ \bibinfo
  {pages} {16048} (\bibinfo {year} {2016})}\BibitemShut {NoStop}%
\bibitem [{\citenamefont {Koshelev}\ \emph {et~al.}(2019)\citenamefont
  {Koshelev}, \citenamefont {Bogdanov},\ and\ \citenamefont
  {Kivshar}}]{koshelev2019}%
  \BibitemOpen
  \bibfield  {author} {\bibinfo {author} {\bibfnamefont {K.}~\bibnamefont
  {Koshelev}}, \bibinfo {author} {\bibfnamefont {A.}~\bibnamefont {Bogdanov}},\
  and\ \bibinfo {author} {\bibfnamefont {Y.}~\bibnamefont {Kivshar}},\
  }\bibfield  {title} {\bibinfo {title} {Meta-optics and bound states in the
  continuum},\ }\href {https://doi.org/10.1016/j.scib.2018.12.003} {\bibfield
  {journal} {\bibinfo  {journal} {Sci. Bull.}\ }\textbf {\bibinfo {volume}
  {64}},\ \bibinfo {pages} {836} (\bibinfo {year} {2019})}\BibitemShut
  {NoStop}%
\bibitem [{\citenamefont {Tonkaev}\ and\ \citenamefont
  {Kivshar}(2020)}]{tonkaev2020}%
  \BibitemOpen
  \bibfield  {author} {\bibinfo {author} {\bibfnamefont {P.}~\bibnamefont
  {Tonkaev}}\ and\ \bibinfo {author} {\bibfnamefont {Y.}~\bibnamefont
  {Kivshar}},\ }\bibfield  {title} {\bibinfo {title} {High-{Q} dielectric
  {M}ie-resonant nanostructures},\ }\href
  {https://doi.org/10.1134/S0021364020220038} {\bibfield  {journal} {\bibinfo
  {journal} {JETP Lett.}\ }\textbf {\bibinfo {volume} {112}},\ \bibinfo {pages}
  {615} (\bibinfo {year} {2020})}\BibitemShut {NoStop}%
\bibitem [{\citenamefont {Lee}\ \emph {et~al.}(2012)\citenamefont {Lee},
  \citenamefont {Zhen}, \citenamefont {Chua}, \citenamefont {Qiu},
  \citenamefont {Joannopoulos}, \citenamefont {Solja{\v{c}}i{\'c}},\ and\
  \citenamefont {Shapira}}]{lee2012}%
  \BibitemOpen
  \bibfield  {author} {\bibinfo {author} {\bibfnamefont {J.}~\bibnamefont
  {Lee}}, \bibinfo {author} {\bibfnamefont {B.}~\bibnamefont {Zhen}}, \bibinfo
  {author} {\bibfnamefont {S.-L.}\ \bibnamefont {Chua}}, \bibinfo {author}
  {\bibfnamefont {W.}~\bibnamefont {Qiu}}, \bibinfo {author} {\bibfnamefont
  {J.~D.}\ \bibnamefont {Joannopoulos}}, \bibinfo {author} {\bibfnamefont
  {M.}~\bibnamefont {Solja{\v{c}}i{\'c}}},\ and\ \bibinfo {author}
  {\bibfnamefont {O.}~\bibnamefont {Shapira}},\ }\bibfield  {title} {\bibinfo
  {title} {Observation and differentiation of unique high-{Q} optical
  resonances near zero wave vector in macroscopic photonic crystal slabs},\
  }\href {https://doi.org/10.1103/PhysRevLett.109.067401} {\bibfield  {journal}
  {\bibinfo  {journal} {Phys. Rev. Lett.}\ }\textbf {\bibinfo {volume} {109}},\
  \bibinfo {pages} {067401} (\bibinfo {year} {2012})}\BibitemShut {NoStop}%
\bibitem [{\citenamefont {Friedrich}\ and\ \citenamefont
  {Wintgen}(1985)}]{Friedrich1985}%
  \BibitemOpen
  \bibfield  {author} {\bibinfo {author} {\bibfnamefont {H.}~\bibnamefont
  {Friedrich}}\ and\ \bibinfo {author} {\bibfnamefont {D.}~\bibnamefont
  {Wintgen}},\ }\bibfield  {title} {\bibinfo {title} {Interfering resonances
  and bound states in the continuum},\ }\href
  {https://doi.org/10.1103/PhysRevA.32.3231} {\bibfield  {journal} {\bibinfo
  {journal} {Phys. Rev. A}\ }\textbf {\bibinfo {volume} {32}},\ \bibinfo
  {pages} {3231} (\bibinfo {year} {1985})}\BibitemShut {NoStop}%
\bibitem [{\citenamefont {Hsu}\ \emph {et~al.}(2013)\citenamefont {Hsu},
  \citenamefont {Zhen}, \citenamefont {Lee}, \citenamefont {Chua},
  \citenamefont {Johnson}, \citenamefont {Joannopoulos},\ and\ \citenamefont
  {Solja{\v{c}}i{\'c}}}]{hsu2013}%
  \BibitemOpen
  \bibfield  {author} {\bibinfo {author} {\bibfnamefont {C.~W.}\ \bibnamefont
  {Hsu}}, \bibinfo {author} {\bibfnamefont {B.}~\bibnamefont {Zhen}}, \bibinfo
  {author} {\bibfnamefont {J.}~\bibnamefont {Lee}}, \bibinfo {author}
  {\bibfnamefont {S.-L.}\ \bibnamefont {Chua}}, \bibinfo {author}
  {\bibfnamefont {S.~G.}\ \bibnamefont {Johnson}}, \bibinfo {author}
  {\bibfnamefont {J.~D.}\ \bibnamefont {Joannopoulos}},\ and\ \bibinfo {author}
  {\bibfnamefont {M.}~\bibnamefont {Solja{\v{c}}i{\'c}}},\ }\bibfield  {title}
  {\bibinfo {title} {Observation of trapped light within the radiation
  continuum},\ }\href {https://doi.org/10.1038/nature12289} {\bibfield
  {journal} {\bibinfo  {journal} {Nature}\ }\textbf {\bibinfo {volume} {499}},\
  \bibinfo {pages} {188} (\bibinfo {year} {2013})}\BibitemShut {NoStop}%
\bibitem [{\citenamefont {Monticone}\ and\ \citenamefont
  {Alu}(2014)}]{monticone2014}%
  \BibitemOpen
  \bibfield  {author} {\bibinfo {author} {\bibfnamefont {F.}~\bibnamefont
  {Monticone}}\ and\ \bibinfo {author} {\bibfnamefont {A.}~\bibnamefont
  {Alu}},\ }\bibfield  {title} {\bibinfo {title} {Embedded photonic eigenvalues
  in {3D} nanostructures},\ }\href
  {https://doi.org/10.1103/PhysRevLett.112.213903} {\bibfield  {journal}
  {\bibinfo  {journal} {Phys. Rev. Lett.}\ }\textbf {\bibinfo {volume} {112}},\
  \bibinfo {pages} {213903} (\bibinfo {year} {2014})}\BibitemShut {NoStop}%
\bibitem [{\citenamefont {Wiersig}(2006)}]{wiersig2006}%
  \BibitemOpen
  \bibfield  {author} {\bibinfo {author} {\bibfnamefont {J.}~\bibnamefont
  {Wiersig}},\ }\bibfield  {title} {\bibinfo {title} {Formation of long-lived,
  scarlike modes near avoided resonance crossings in optical microcavities},\
  }\href {https://doi.org/10.1103/PhysRevLett.97.253901} {\bibfield  {journal}
  {\bibinfo  {journal} {Phys. Rev. Lett.}\ }\textbf {\bibinfo {volume} {97}},\
  \bibinfo {pages} {253901} (\bibinfo {year} {2006})}\BibitemShut {NoStop}%
\bibitem [{\citenamefont {Song}\ and\ \citenamefont {Cao}(2010)}]{song2010}%
  \BibitemOpen
  \bibfield  {author} {\bibinfo {author} {\bibfnamefont {Q.~H.}\ \bibnamefont
  {Song}}\ and\ \bibinfo {author} {\bibfnamefont {H.}~\bibnamefont {Cao}},\
  }\bibfield  {title} {\bibinfo {title} {Improving optical confinement in
  nanostructures via external mode coupling},\ }\href
  {https://doi.org/10.1103/PhysRevLett.105.053902} {\bibfield  {journal}
  {\bibinfo  {journal} {Phys. Rev. Lett.}\ }\textbf {\bibinfo {volume} {105}},\
  \bibinfo {pages} {053902} (\bibinfo {year} {2010})}\BibitemShut {NoStop}%
\bibitem [{\citenamefont {Rybin}\ \emph {et~al.}(2017)\citenamefont {Rybin},
  \citenamefont {Koshelev}, \citenamefont {Sadrieva}, \citenamefont {Samusev},
  \citenamefont {Bogdanov}, \citenamefont {Limonov},\ and\ \citenamefont
  {Kivshar}}]{rybin2017}%
  \BibitemOpen
  \bibfield  {author} {\bibinfo {author} {\bibfnamefont {M.~V.}\ \bibnamefont
  {Rybin}}, \bibinfo {author} {\bibfnamefont {K.~L.}\ \bibnamefont {Koshelev}},
  \bibinfo {author} {\bibfnamefont {Z.~F.}\ \bibnamefont {Sadrieva}}, \bibinfo
  {author} {\bibfnamefont {K.~B.}\ \bibnamefont {Samusev}}, \bibinfo {author}
  {\bibfnamefont {A.~A.}\ \bibnamefont {Bogdanov}}, \bibinfo {author}
  {\bibfnamefont {M.~F.}\ \bibnamefont {Limonov}},\ and\ \bibinfo {author}
  {\bibfnamefont {Y.~S.}\ \bibnamefont {Kivshar}},\ }\bibfield  {title}
  {\bibinfo {title} {High-{Q} supercavity modes in subwavelength dielectric
  resonators},\ }\href {https://doi.org/10.1103/PhysRevLett.119.243901}
  {\bibfield  {journal} {\bibinfo  {journal} {Phys. Rev. Lett.}\ }\textbf
  {\bibinfo {volume} {119}},\ \bibinfo {pages} {243901} (\bibinfo {year}
  {2017})}\BibitemShut {NoStop}%
\bibitem [{\citenamefont {Bogdanov}\ \emph {et~al.}(2019)\citenamefont
  {Bogdanov}, \citenamefont {Koshelev}, \citenamefont {Kapitanova},
  \citenamefont {Rybin}, \citenamefont {Gladyshev}, \citenamefont {Sadrieva},
  \citenamefont {Samusev}, \citenamefont {Kivshar},\ and\ \citenamefont
  {Limonov}}]{bogdanov2019}%
  \BibitemOpen
  \bibfield  {author} {\bibinfo {author} {\bibfnamefont {A.~A.}\ \bibnamefont
  {Bogdanov}}, \bibinfo {author} {\bibfnamefont {K.~L.}\ \bibnamefont
  {Koshelev}}, \bibinfo {author} {\bibfnamefont {P.~V.}\ \bibnamefont
  {Kapitanova}}, \bibinfo {author} {\bibfnamefont {M.~V.}\ \bibnamefont
  {Rybin}}, \bibinfo {author} {\bibfnamefont {S.~A.}\ \bibnamefont
  {Gladyshev}}, \bibinfo {author} {\bibfnamefont {Z.~F.}\ \bibnamefont
  {Sadrieva}}, \bibinfo {author} {\bibfnamefont {K.~B.}\ \bibnamefont
  {Samusev}}, \bibinfo {author} {\bibfnamefont {Y.~S.}\ \bibnamefont
  {Kivshar}},\ and\ \bibinfo {author} {\bibfnamefont {M.~F.}\ \bibnamefont
  {Limonov}},\ }\bibfield  {title} {\bibinfo {title} {Bound states in the
  continuum and {F}ano resonances in the strong mode coupling regime},\ }\href
  {https://doi.org/10.1117/1.AP.1.1.016001} {\bibfield  {journal} {\bibinfo
  {journal} {Adv. Photonics}\ }\textbf {\bibinfo {volume} {1}},\ \bibinfo
  {pages} {016001} (\bibinfo {year} {2019})}\BibitemShut {NoStop}%
\bibitem [{\citenamefont {Koshelev}\ and\ \citenamefont
  {Kivshar}(2020)}]{koshelev2020c}%
  \BibitemOpen
  \bibfield  {author} {\bibinfo {author} {\bibfnamefont {K.}~\bibnamefont
  {Koshelev}}\ and\ \bibinfo {author} {\bibfnamefont {Y.}~\bibnamefont
  {Kivshar}},\ }\bibfield  {title} {\bibinfo {title} {Dielectric resonant
  metaphotonics},\ }\href {https://doi.org/10.1021/acsphotonics.0c01315}
  {\bibfield  {journal} {\bibinfo  {journal} {ACS Photonics}\ }\textbf
  {\bibinfo {volume} {8}},\ \bibinfo {pages} {102} (\bibinfo {year}
  {2020})}\BibitemShut {NoStop}%
\bibitem [{\citenamefont {Melik-Gaykazyan}\ \emph {et~al.}(2021)\citenamefont
  {Melik-Gaykazyan}, \citenamefont {Koshelev}, \citenamefont {Choi},
  \citenamefont {Kruk}, \citenamefont {Bogdanov}, \citenamefont {Park},\ and\
  \citenamefont {Kivshar}}]{melik2021}%
  \BibitemOpen
  \bibfield  {author} {\bibinfo {author} {\bibfnamefont {E.}~\bibnamefont
  {Melik-Gaykazyan}}, \bibinfo {author} {\bibfnamefont {K.}~\bibnamefont
  {Koshelev}}, \bibinfo {author} {\bibfnamefont {J.-H.}\ \bibnamefont {Choi}},
  \bibinfo {author} {\bibfnamefont {S.~S.}\ \bibnamefont {Kruk}}, \bibinfo
  {author} {\bibfnamefont {A.}~\bibnamefont {Bogdanov}}, \bibinfo {author}
  {\bibfnamefont {H.-G.}\ \bibnamefont {Park}},\ and\ \bibinfo {author}
  {\bibfnamefont {Y.}~\bibnamefont {Kivshar}},\ }\bibfield  {title} {\bibinfo
  {title} {From {F}ano to quasi-{BIC} resonances in individual dielectric
  nanoantennas},\ }\href {https://doi.org/10.1021/acs.nanolett.0c04660}
  {\bibfield  {journal} {\bibinfo  {journal} {Nano Lett.}\ }\textbf {\bibinfo
  {volume} {21}},\ \bibinfo {pages} {1765} (\bibinfo {year}
  {2021})}\BibitemShut {NoStop}%
\bibitem [{\citenamefont {Koshelev}\ \emph
  {et~al.}(2020{\natexlab{a}})\citenamefont {Koshelev}, \citenamefont
  {Bogdanov},\ and\ \citenamefont {Kivshar}}]{koshelev2020b}%
  \BibitemOpen
  \bibfield  {author} {\bibinfo {author} {\bibfnamefont {K.}~\bibnamefont
  {Koshelev}}, \bibinfo {author} {\bibfnamefont {A.}~\bibnamefont {Bogdanov}},\
  and\ \bibinfo {author} {\bibfnamefont {Y.}~\bibnamefont {Kivshar}},\
  }\bibfield  {title} {\bibinfo {title} {Engineering with bound states in the
  continuum},\ }\href {https://doi.org/10.1364/OPN.31.1.000038} {\bibfield
  {journal} {\bibinfo  {journal} {Opt. Photonics News}\ }\textbf {\bibinfo
  {volume} {31}},\ \bibinfo {pages} {38} (\bibinfo {year}
  {2020}{\natexlab{a}})}\BibitemShut {NoStop}%
\bibitem [{\citenamefont {Carletti}\ \emph {et~al.}(2018)\citenamefont
  {Carletti}, \citenamefont {Koshelev}, \citenamefont {De~Angelis},\ and\
  \citenamefont {Kivshar}}]{carletti2018}%
  \BibitemOpen
  \bibfield  {author} {\bibinfo {author} {\bibfnamefont {L.}~\bibnamefont
  {Carletti}}, \bibinfo {author} {\bibfnamefont {K.}~\bibnamefont {Koshelev}},
  \bibinfo {author} {\bibfnamefont {C.}~\bibnamefont {De~Angelis}},\ and\
  \bibinfo {author} {\bibfnamefont {Y.}~\bibnamefont {Kivshar}},\ }\bibfield
  {title} {\bibinfo {title} {Giant nonlinear response at the nanoscale driven
  by bound states in the continuum},\ }\href
  {https://doi.org/10.1103/PhysRevLett.121.033903} {\bibfield  {journal}
  {\bibinfo  {journal} {Phys. Rev. Lett.}\ }\textbf {\bibinfo {volume} {121}},\
  \bibinfo {pages} {033903} (\bibinfo {year} {2018})}\BibitemShut {NoStop}%
\bibitem [{\citenamefont {Carletti}\ \emph {et~al.}(2019)\citenamefont
  {Carletti}, \citenamefont {Kruk}, \citenamefont {Bogdanov}, \citenamefont
  {De~Angelis},\ and\ \citenamefont {Kivshar}}]{carletti2019}%
  \BibitemOpen
  \bibfield  {author} {\bibinfo {author} {\bibfnamefont {L.}~\bibnamefont
  {Carletti}}, \bibinfo {author} {\bibfnamefont {S.~S.}\ \bibnamefont {Kruk}},
  \bibinfo {author} {\bibfnamefont {A.~A.}\ \bibnamefont {Bogdanov}}, \bibinfo
  {author} {\bibfnamefont {C.}~\bibnamefont {De~Angelis}},\ and\ \bibinfo
  {author} {\bibfnamefont {Y.}~\bibnamefont {Kivshar}},\ }\bibfield  {title}
  {\bibinfo {title} {High-harmonic generation at the nanoscale boosted by bound
  states in the continuum},\ }\href
  {https://doi.org/10.1103/PhysRevResearch.1.023016} {\bibfield  {journal}
  {\bibinfo  {journal} {Phys. Rev. Res.}\ }\textbf {\bibinfo {volume} {1}},\
  \bibinfo {pages} {023016} (\bibinfo {year} {2019})}\BibitemShut {NoStop}%
\bibitem [{\citenamefont {Koshelev}\ \emph
  {et~al.}(2020{\natexlab{b}})\citenamefont {Koshelev}, \citenamefont {Kruk},
  \citenamefont {Melik-Gaykazyan}, \citenamefont {Choi}, \citenamefont
  {Bogdanov}, \citenamefont {Park},\ and\ \citenamefont
  {Kivshar}}]{koshelev2020}%
  \BibitemOpen
  \bibfield  {author} {\bibinfo {author} {\bibfnamefont {K.}~\bibnamefont
  {Koshelev}}, \bibinfo {author} {\bibfnamefont {S.}~\bibnamefont {Kruk}},
  \bibinfo {author} {\bibfnamefont {E.}~\bibnamefont {Melik-Gaykazyan}},
  \bibinfo {author} {\bibfnamefont {J.-H.}\ \bibnamefont {Choi}}, \bibinfo
  {author} {\bibfnamefont {A.}~\bibnamefont {Bogdanov}}, \bibinfo {author}
  {\bibfnamefont {H.-G.}\ \bibnamefont {Park}},\ and\ \bibinfo {author}
  {\bibfnamefont {Y.}~\bibnamefont {Kivshar}},\ }\bibfield  {title} {\bibinfo
  {title} {Subwavelength dielectric resonators for nonlinear nanophotonics},\
  }\href {https://doi.org/10.1126/science.aaz3985} {\bibfield  {journal}
  {\bibinfo  {journal} {Science}\ }\textbf {\bibinfo {volume} {367}},\ \bibinfo
  {pages} {288} (\bibinfo {year} {2020}{\natexlab{b}})}\BibitemShut {NoStop}%
\bibitem [{\citenamefont {Mylnikov}\ \emph {et~al.}(2020)\citenamefont
  {Mylnikov}, \citenamefont {Ha}, \citenamefont {Pan}, \citenamefont
  {Valuckas}, \citenamefont {Paniagua-Dom\'{i}nguez}, \citenamefont {Demir},\
  and\ \citenamefont {Kuznetsov}}]{mylnikov2020}%
  \BibitemOpen
  \bibfield  {author} {\bibinfo {author} {\bibfnamefont {V.}~\bibnamefont
  {Mylnikov}}, \bibinfo {author} {\bibfnamefont {S.~T.}\ \bibnamefont {Ha}},
  \bibinfo {author} {\bibfnamefont {Z.}~\bibnamefont {Pan}}, \bibinfo {author}
  {\bibfnamefont {V.}~\bibnamefont {Valuckas}}, \bibinfo {author}
  {\bibfnamefont {R.}~\bibnamefont {Paniagua-Dom\'{i}nguez}}, \bibinfo {author}
  {\bibfnamefont {H.~V.}\ \bibnamefont {Demir}},\ and\ \bibinfo {author}
  {\bibfnamefont {A.~I.}\ \bibnamefont {Kuznetsov}},\ }\bibfield  {title}
  {\bibinfo {title} {Lasing action in single subwavelength particles supporting
  supercavity modes},\ }\href {https://doi.org/10.1021/acsnano.0c02730}
  {\bibfield  {journal} {\bibinfo  {journal} {ACS Nano}\ }\textbf {\bibinfo
  {volume} {14}},\ \bibinfo {pages} {7338} (\bibinfo {year}
  {2020})}\BibitemShut {NoStop}%
\bibitem [{\citenamefont {Lalanne}\ \emph {et~al.}(2018)\citenamefont
  {Lalanne}, \citenamefont {Yan}, \citenamefont {Vynck}, \citenamefont
  {Sauvan},\ and\ \citenamefont {Hugonin}}]{lalanne2018}%
  \BibitemOpen
  \bibfield  {author} {\bibinfo {author} {\bibfnamefont {P.}~\bibnamefont
  {Lalanne}}, \bibinfo {author} {\bibfnamefont {W.}~\bibnamefont {Yan}},
  \bibinfo {author} {\bibfnamefont {K.}~\bibnamefont {Vynck}}, \bibinfo
  {author} {\bibfnamefont {C.}~\bibnamefont {Sauvan}},\ and\ \bibinfo {author}
  {\bibfnamefont {J.-P.}\ \bibnamefont {Hugonin}},\ }\bibfield  {title}
  {\bibinfo {title} {Light interaction with photonic and plasmonic
  resonances},\ }\href {https://doi.org/10.1002/lpor.201700113} {\bibfield
  {journal} {\bibinfo  {journal} {Laser Photonics Rev.}\ }\textbf {\bibinfo
  {volume} {12}},\ \bibinfo {pages} {1700113} (\bibinfo {year}
  {2018})}\BibitemShut {NoStop}%
\bibitem [{\citenamefont {Koshelev}\ \emph {et~al.}(2018)\citenamefont
  {Koshelev}, \citenamefont {Lepeshov}, \citenamefont {Liu}, \citenamefont
  {Bogdanov},\ and\ \citenamefont {Kivshar}}]{koshelev2018}%
  \BibitemOpen
  \bibfield  {author} {\bibinfo {author} {\bibfnamefont {K.}~\bibnamefont
  {Koshelev}}, \bibinfo {author} {\bibfnamefont {S.}~\bibnamefont {Lepeshov}},
  \bibinfo {author} {\bibfnamefont {M.}~\bibnamefont {Liu}}, \bibinfo {author}
  {\bibfnamefont {A.}~\bibnamefont {Bogdanov}},\ and\ \bibinfo {author}
  {\bibfnamefont {Y.}~\bibnamefont {Kivshar}},\ }\bibfield  {title} {\bibinfo
  {title} {Asymmetric metasurfaces with high-{Q} resonances governed by bound
  states in the continuum},\ }\href
  {https://doi.org/10.1103/PhysRevLett.121.193903} {\bibfield  {journal}
  {\bibinfo  {journal} {Phys. Rev. Lett.}\ }\textbf {\bibinfo {volume} {121}},\
  \bibinfo {pages} {193903} (\bibinfo {year} {2018})}\BibitemShut {NoStop}%
\bibitem [{\citenamefont {Kristensen}\ \emph {et~al.}(2020)\citenamefont
  {Kristensen}, \citenamefont {Herrmann}, \citenamefont {Intravaia},\ and\
  \citenamefont {Busch}}]{kristensen2020}%
  \BibitemOpen
  \bibfield  {author} {\bibinfo {author} {\bibfnamefont {P.~T.}\ \bibnamefont
  {Kristensen}}, \bibinfo {author} {\bibfnamefont {K.}~\bibnamefont
  {Herrmann}}, \bibinfo {author} {\bibfnamefont {F.}~\bibnamefont
  {Intravaia}},\ and\ \bibinfo {author} {\bibfnamefont {K.}~\bibnamefont
  {Busch}},\ }\bibfield  {title} {\bibinfo {title} {Modeling electromagnetic
  resonators using quasinormal modes},\ }\href
  {https://doi.org/10.1364/AOP.377940} {\bibfield  {journal} {\bibinfo
  {journal} {Adv. Opt. Photonics}\ }\textbf {\bibinfo {volume} {12}},\ \bibinfo
  {pages} {612} (\bibinfo {year} {2020})}\BibitemShut {NoStop}%
\bibitem [{\citenamefont {Wu}\ \emph {et~al.}(2021)\citenamefont {Wu},
  \citenamefont {Gurioli},\ and\ \citenamefont {Lalanne}}]{wu2021}%
  \BibitemOpen
  \bibfield  {author} {\bibinfo {author} {\bibfnamefont {T.}~\bibnamefont
  {Wu}}, \bibinfo {author} {\bibfnamefont {M.}~\bibnamefont {Gurioli}},\ and\
  \bibinfo {author} {\bibfnamefont {P.}~\bibnamefont {Lalanne}},\ }\bibfield
  {title} {\bibinfo {title} {Nanoscale light confinement: the {Q}’s and
  {V}’s},\ }\href {https://doi.org/10.1021/acsphotonics.1c00336} {\bibfield
  {journal} {\bibinfo  {journal} {ACS Photonics}\ }\textbf {\bibinfo {volume}
  {8}},\ \bibinfo {pages} {1522} (\bibinfo {year} {2021})}\BibitemShut
  {NoStop}%
\bibitem [{\citenamefont {Sauvan}\ \emph {et~al.}(2013)\citenamefont {Sauvan},
  \citenamefont {Hugonin}, \citenamefont {Maksymov},\ and\ \citenamefont
  {Lalanne}}]{sauvan2013}%
  \BibitemOpen
  \bibfield  {author} {\bibinfo {author} {\bibfnamefont {C.}~\bibnamefont
  {Sauvan}}, \bibinfo {author} {\bibfnamefont {J.-P.}\ \bibnamefont {Hugonin}},
  \bibinfo {author} {\bibfnamefont {I.~S.}\ \bibnamefont {Maksymov}},\ and\
  \bibinfo {author} {\bibfnamefont {P.}~\bibnamefont {Lalanne}},\ }\bibfield
  {title} {\bibinfo {title} {Theory of the spontaneous optical emission of
  nanosize photonic and plasmon resonators},\ }\href
  {https://doi.org/10.1103/PhysRevLett.110.237401} {\bibfield  {journal}
  {\bibinfo  {journal} {Phys. Rev. Lett.}\ }\textbf {\bibinfo {volume} {110}},\
  \bibinfo {pages} {237401} (\bibinfo {year} {2013})}\BibitemShut {NoStop}%
\bibitem [{\citenamefont {Ge}\ \emph {et~al.}(2014)\citenamefont {Ge},
  \citenamefont {Kristensen}, \citenamefont {Young},\ and\ \citenamefont
  {Hughes}}]{ge2014}%
  \BibitemOpen
  \bibfield  {author} {\bibinfo {author} {\bibfnamefont {R.-C.}\ \bibnamefont
  {Ge}}, \bibinfo {author} {\bibfnamefont {P.~T.}\ \bibnamefont {Kristensen}},
  \bibinfo {author} {\bibfnamefont {J.~F.}\ \bibnamefont {Young}},\ and\
  \bibinfo {author} {\bibfnamefont {S.}~\bibnamefont {Hughes}},\ }\bibfield
  {title} {\bibinfo {title} {Quasinormal mode approach to modelling
  light-emission and propagation in nanoplasmonics},\ }\href
  {https://doi.org/10.1088/1367-2630/16/11/113048} {\bibfield  {journal}
  {\bibinfo  {journal} {New J. Phys.}\ }\textbf {\bibinfo {volume} {16}},\
  \bibinfo {pages} {113048} (\bibinfo {year} {2014})}\BibitemShut {NoStop}%
\bibitem [{\citenamefont {Muljarov}\ and\ \citenamefont
  {Langbein}(2016)}]{muljarov2016}%
  \BibitemOpen
  \bibfield  {author} {\bibinfo {author} {\bibfnamefont {E.~A.}\ \bibnamefont
  {Muljarov}}\ and\ \bibinfo {author} {\bibfnamefont {W.}~\bibnamefont
  {Langbein}},\ }\bibfield  {title} {\bibinfo {title} {Exact mode volume and
  {P}urcell factor of open optical systems},\ }\href
  {https://doi.org/10.1103/PhysRevB.94.235438} {\bibfield  {journal} {\bibinfo
  {journal} {Phys. Rev. B}\ }\textbf {\bibinfo {volume} {94}},\ \bibinfo
  {pages} {235438} (\bibinfo {year} {2016})}\BibitemShut {NoStop}%
\bibitem [{\citenamefont {Zschiedrich}\ \emph {et~al.}(2018)\citenamefont
  {Zschiedrich}, \citenamefont {Binkowski}, \citenamefont {Nikolay},
  \citenamefont {Benson}, \citenamefont {Kewes},\ and\ \citenamefont
  {Burger}}]{Zschiedrich2018}%
  \BibitemOpen
  \bibfield  {author} {\bibinfo {author} {\bibfnamefont {L.}~\bibnamefont
  {Zschiedrich}}, \bibinfo {author} {\bibfnamefont {F.}~\bibnamefont
  {Binkowski}}, \bibinfo {author} {\bibfnamefont {N.}~\bibnamefont {Nikolay}},
  \bibinfo {author} {\bibfnamefont {O.}~\bibnamefont {Benson}}, \bibinfo
  {author} {\bibfnamefont {G.}~\bibnamefont {Kewes}},\ and\ \bibinfo {author}
  {\bibfnamefont {S.}~\bibnamefont {Burger}},\ }\bibfield  {title} {\bibinfo
  {title} {Riesz-projection-based theory of light-matter interaction in
  dispersive nanoresonators},\ }\href
  {https://doi.org/10.1103/PhysRevA.98.043806} {\bibfield  {journal} {\bibinfo
  {journal} {Phys. Rev. A}\ }\textbf {\bibinfo {volume} {98}},\ \bibinfo
  {pages} {043806} (\bibinfo {year} {2018})}\BibitemShut {NoStop}%
\bibitem [{\citenamefont {Kaganskiy}\ \emph {et~al.}(2018)\citenamefont
  {Kaganskiy}, \citenamefont {Gericke}, \citenamefont {Heuser}, \citenamefont
  {Heindel}, \citenamefont {Porte},\ and\ \citenamefont
  {Reitzenstein}}]{Kaganskiy_2018}%
  \BibitemOpen
  \bibfield  {author} {\bibinfo {author} {\bibfnamefont {A.}~\bibnamefont
  {Kaganskiy}}, \bibinfo {author} {\bibfnamefont {F.}~\bibnamefont {Gericke}},
  \bibinfo {author} {\bibfnamefont {T.}~\bibnamefont {Heuser}}, \bibinfo
  {author} {\bibfnamefont {T.}~\bibnamefont {Heindel}}, \bibinfo {author}
  {\bibfnamefont {X.}~\bibnamefont {Porte}},\ and\ \bibinfo {author}
  {\bibfnamefont {S.}~\bibnamefont {Reitzenstein}},\ }\bibfield  {title}
  {\bibinfo {title} {Micropillars with a controlled number of site-controlled
  quantum dots},\ }\href {https://doi.org/10.1063/1.5017692} {\bibfield
  {journal} {\bibinfo  {journal} {Appl. Phys. Lett.}\ }\textbf {\bibinfo
  {volume} {112}},\ \bibinfo {pages} {071101} (\bibinfo {year}
  {2018})}\BibitemShut {NoStop}%
\bibitem [{\citenamefont {Yi}\ \emph {et~al.}(2019)\citenamefont {Yi},
  \citenamefont {Kullig}, \citenamefont {Hentschel},\ and\ \citenamefont
  {Wiersig}}]{yi2019}%
  \BibitemOpen
  \bibfield  {author} {\bibinfo {author} {\bibfnamefont {C.-H.}\ \bibnamefont
  {Yi}}, \bibinfo {author} {\bibfnamefont {J.}~\bibnamefont {Kullig}}, \bibinfo
  {author} {\bibfnamefont {M.}~\bibnamefont {Hentschel}},\ and\ \bibinfo
  {author} {\bibfnamefont {J.}~\bibnamefont {Wiersig}},\ }\bibfield  {title}
  {\bibinfo {title} {Non-{H}ermitian degeneracies of internal--external mode
  pairs in dielectric microdisks},\ }\href
  {https://doi.org/10.1364/PRJ.7.000464} {\bibfield  {journal} {\bibinfo
  {journal} {Photonics Res.}\ }\textbf {\bibinfo {volume} {7}},\ \bibinfo
  {pages} {464} (\bibinfo {year} {2019})}\BibitemShut {NoStop}%
\bibitem [{\citenamefont {Heiss}(2000)}]{heiss2000}%
  \BibitemOpen
  \bibfield  {author} {\bibinfo {author} {\bibfnamefont {W.~D.}\ \bibnamefont
  {Heiss}},\ }\bibfield  {title} {\bibinfo {title} {Repulsion of resonance
  states and exceptional points},\ }\href
  {https://doi.org/10.1103/PhysRevE.61.929} {\bibfield  {journal} {\bibinfo
  {journal} {Phys. Rev. E}\ }\textbf {\bibinfo {volume} {61}},\ \bibinfo
  {pages} {929} (\bibinfo {year} {2000})}\BibitemShut {NoStop}%
\bibitem [{\citenamefont {Pomplun}\ \emph {et~al.}(2007)\citenamefont
  {Pomplun}, \citenamefont {Burger}, \citenamefont {Zschiedrich},\ and\
  \citenamefont {Schmidt}}]{pomplun2007}%
  \BibitemOpen
  \bibfield  {author} {\bibinfo {author} {\bibfnamefont {J.}~\bibnamefont
  {Pomplun}}, \bibinfo {author} {\bibfnamefont {S.}~\bibnamefont {Burger}},
  \bibinfo {author} {\bibfnamefont {L.}~\bibnamefont {Zschiedrich}},\ and\
  \bibinfo {author} {\bibfnamefont {F.}~\bibnamefont {Schmidt}},\ }\bibfield
  {title} {\bibinfo {title} {Adaptive finite element method for simulation of
  optical nano structures},\ }\href {https://doi.org/10.1002/pssb.200743192}
  {\bibfield  {journal} {\bibinfo  {journal} {Phys. Status Solidi B}\ }\textbf
  {\bibinfo {volume} {244}},\ \bibinfo {pages} {3419} (\bibinfo {year}
  {2007})}\BibitemShut {NoStop}%
\bibitem [{\citenamefont {Huang}\ \emph {et~al.}(2021)\citenamefont {Huang},
  \citenamefont {Xu}, \citenamefont {Rahmani}, \citenamefont {Neshev},\ and\
  \citenamefont {Miroshnichenko}}]{huang2021}%
  \BibitemOpen
  \bibfield  {author} {\bibinfo {author} {\bibfnamefont {L.}~\bibnamefont
  {Huang}}, \bibinfo {author} {\bibfnamefont {L.}~\bibnamefont {Xu}}, \bibinfo
  {author} {\bibfnamefont {M.}~\bibnamefont {Rahmani}}, \bibinfo {author}
  {\bibfnamefont {D.}~\bibnamefont {Neshev}},\ and\ \bibinfo {author}
  {\bibfnamefont {A.~E.}\ \bibnamefont {Miroshnichenko}},\ }\bibfield  {title}
  {\bibinfo {title} {Pushing the limit of high-{Q} mode of a single dielectric
  nanocavity},\ }\href {https://doi.org/10.1117/1.AP.3.1.016004} {\bibfield
  {journal} {\bibinfo  {journal} {Adv. Photonics}\ }\textbf {\bibinfo {volume}
  {3}},\ \bibinfo {pages} {016004} (\bibinfo {year} {2021})}\BibitemShut
  {NoStop}%
\bibitem [{\citenamefont {Yan}\ \emph {et~al.}(2020)\citenamefont {Yan},
  \citenamefont {Lalanne},\ and\ \citenamefont {Qiu}}]{yan2020}%
  \BibitemOpen
  \bibfield  {author} {\bibinfo {author} {\bibfnamefont {W.}~\bibnamefont
  {Yan}}, \bibinfo {author} {\bibfnamefont {P.}~\bibnamefont {Lalanne}},\ and\
  \bibinfo {author} {\bibfnamefont {M.}~\bibnamefont {Qiu}},\ }\bibfield
  {title} {\bibinfo {title} {Shape deformation of nanoresonator: {A}
  quasinormal-mode perturbation theory},\ }\href
  {https://doi.org/10.1103/PhysRevLett.125.013901} {\bibfield  {journal}
  {\bibinfo  {journal} {Phys. Rev. Lett.}\ }\textbf {\bibinfo {volume} {125}},\
  \bibinfo {pages} {013901} (\bibinfo {year} {2020})}\BibitemShut {NoStop}%
\bibitem [{\citenamefont {Heiss}(2012)}]{heiss2012}%
  \BibitemOpen
  \bibfield  {author} {\bibinfo {author} {\bibfnamefont {W.~D.}\ \bibnamefont
  {Heiss}},\ }\bibfield  {title} {\bibinfo {title} {The physics of exceptional
  points},\ }\href {https://doi.org/10.1088/1751-8113/45/44/444016} {\bibfield
  {journal} {\bibinfo  {journal} {J. Phys. A}\ }\textbf {\bibinfo {volume}
  {45}},\ \bibinfo {pages} {444016} (\bibinfo {year} {2012})}\BibitemShut
  {NoStop}%
\bibitem [{\citenamefont {Rodriguez}(2016)}]{rodriguez2016}%
  \BibitemOpen
  \bibfield  {author} {\bibinfo {author} {\bibfnamefont {S.~R.-K.}\
  \bibnamefont {Rodriguez}},\ }\bibfield  {title} {\bibinfo {title} {Classical
  and quantum distinctions between weak and strong coupling},\ }\href
  {https://doi.org/10.1088/0143-0807/37/2/025802} {\bibfield  {journal}
  {\bibinfo  {journal} {Eur. J. Phys.}\ }\textbf {\bibinfo {volume} {37}},\
  \bibinfo {pages} {025802} (\bibinfo {year} {2016})}\BibitemShut {NoStop}%
\bibitem [{\citenamefont {Deng}\ \emph {et~al.}(2022)\citenamefont {Deng},
  \citenamefont {Li}, \citenamefont {Li}, \citenamefont {Li},\ and\
  \citenamefont {Al{\`u}}}]{deng2021}%
  \BibitemOpen
  \bibfield  {author} {\bibinfo {author} {\bibfnamefont {Z.-L.}\ \bibnamefont
  {Deng}}, \bibinfo {author} {\bibfnamefont {F.-J.}\ \bibnamefont {Li}},
  \bibinfo {author} {\bibfnamefont {H.}~\bibnamefont {Li}}, \bibinfo {author}
  {\bibfnamefont {X.}~\bibnamefont {Li}},\ and\ \bibinfo {author}
  {\bibfnamefont {A.}~\bibnamefont {Al{\`u}}},\ }\bibfield  {title} {\bibinfo
  {title} {Extreme diffraction control in metagratings leveraging bound states
  in the continuum and exceptional points},\ }\href
  {https://doi.org/10.1002/lpor.202100617} {\bibfield  {journal} {\bibinfo
  {journal} {Laser Photon. Rev.}\ ,\ \bibinfo {pages} {2100617}} (\bibinfo
  {year} {2022})}\BibitemShut {NoStop}%
\bibitem [{\citenamefont {Abujetas}\ and\ \citenamefont
  {S{\'a}nchez-Gil}(2021)}]{abujetas2021}%
  \BibitemOpen
  \bibfield  {author} {\bibinfo {author} {\bibfnamefont {D.~R.}\ \bibnamefont
  {Abujetas}}\ and\ \bibinfo {author} {\bibfnamefont {J.~A.}\ \bibnamefont
  {S{\'a}nchez-Gil}},\ }\bibfield  {title} {\bibinfo {title} {Near-field
  excitation of bound states in the continuum in all-dielectric metasurfaces
  through a coupled electric/magnetic dipole model},\ }\href
  {https://doi.org/10.3390/nano11040998} {\bibfield  {journal} {\bibinfo
  {journal} {Nanomaterials}\ }\textbf {\bibinfo {volume} {11}},\ \bibinfo
  {pages} {998} (\bibinfo {year} {2021})}\BibitemShut {NoStop}%
\bibitem [{\citenamefont {Binkowski}\ \emph {et~al.}(2020)\citenamefont
  {Binkowski}, \citenamefont {Betz}, \citenamefont {Colom}, \citenamefont
  {Hammerschmidt}, \citenamefont {Zschiedrich},\ and\ \citenamefont
  {Burger}}]{Binkowski2020}%
  \BibitemOpen
  \bibfield  {author} {\bibinfo {author} {\bibfnamefont {F.}~\bibnamefont
  {Binkowski}}, \bibinfo {author} {\bibfnamefont {F.}~\bibnamefont {Betz}},
  \bibinfo {author} {\bibfnamefont {R.}~\bibnamefont {Colom}}, \bibinfo
  {author} {\bibfnamefont {M.}~\bibnamefont {Hammerschmidt}}, \bibinfo {author}
  {\bibfnamefont {L.}~\bibnamefont {Zschiedrich}},\ and\ \bibinfo {author}
  {\bibfnamefont {S.}~\bibnamefont {Burger}},\ }\bibfield  {title} {\bibinfo
  {title} {Quasinormal mode expansion of optical far-field quantities},\ }\href
  {https://doi.org/10.1103/PhysRevB.102.035432} {\bibfield  {journal} {\bibinfo
   {journal} {Phys. Rev. B}\ }\textbf {\bibinfo {volume} {102}},\ \bibinfo
  {pages} {035432} (\bibinfo {year} {2020})}\BibitemShut {NoStop}%
\bibitem [{\citenamefont {Betz}\ \emph {et~al.}(2021)\citenamefont {Betz},
  \citenamefont {Binkowski},\ and\ \citenamefont {Burger}}]{betz2021}%
  \BibitemOpen
  \bibfield  {author} {\bibinfo {author} {\bibfnamefont {F.}~\bibnamefont
  {Betz}}, \bibinfo {author} {\bibfnamefont {F.}~\bibnamefont {Binkowski}},\
  and\ \bibinfo {author} {\bibfnamefont {S.}~\bibnamefont {Burger}},\
  }\bibfield  {title} {\bibinfo {title} {{\sc {RPE}xpand}: {S}oftware for {R}iesz
  projection expansion of resonance phenomena},\ }\href
  {https://doi.org/10.1016/j.softx.2021.100763} {\bibfield  {journal} {\bibinfo
   {journal} {SoftwareX}\ }\textbf {\bibinfo {volume} {15}},\ \bibinfo {pages}
  {100763} (\bibinfo {year} {2021})}\BibitemShut {NoStop}%
\bibitem [{\citenamefont {Colom}\ \emph {et~al.}(2022)\citenamefont {Colom},
  \citenamefont {Binkowski}, \citenamefont {Betz}, \citenamefont {Kivshar},\
  and\ \citenamefont {Burger}}]{colom_binkowski_betz_data}%
  \BibitemOpen
  \bibfield  {author} {\bibinfo {author} {\bibfnamefont {R.}~\bibnamefont
  {Colom}}, \bibinfo {author} {\bibfnamefont {F.}~\bibnamefont {Binkowski}},
  \bibinfo {author} {\bibfnamefont {F.}~\bibnamefont {Betz}}, \bibinfo {author}
  {\bibfnamefont {Y.}~\bibnamefont {Kivshar}},\ and\ \bibinfo {author}
  {\bibfnamefont {S.}~\bibnamefont {Burger}},\ }\bibfield  {title} {\bibinfo
  {title} {{Source code and simulation results for nanoantennas supporting an
      enhanced Purcell factor due to interfering resonances}},\ }
             \bibinfo {note} {[Data set], Zenodo}
     \href {https://doi.org/10.5281/zenodo.6565850} {https://doi.org/10.5281/zenodo.6565850},\
             (\bibinfo {year} {2022})
             \BibitemShut {NoStop}%
\end{thebibliography}


%


\end{document}